\documentclass{jfm}

\usepackage{subcaption}

\usepackage{graphicx}

\usepackage{natbib,amsmath}
\usepackage{mathtools,bbm}
\usepackage[dvipsnames]{xcolor}
\usepackage{hyperref}

\definecolor{light-gray}{gray}{0.95}
\ifCUPmtlplainloaded \else
  \checkfont{eurm10}
  \iffontfound
    \IfFileExists{upmath.sty}
      {\typeout{^^JFound AMS Euler Roman fonts on the system,
                   using the 'upmath' package.^^J}%
       \usepackage{upmath}}
      {\typeout{^^JFound AMS Euler Roman fonts on the system, but you
                   dont seem to have the}%
       \typeout{'upmath' package installed. JFM.cls can take advantage
                 of these fonts,^^Jif you use 'upmath' package.^^J}%
      }
  \else
  \fi
\fi

\ifCUPmtlplainloaded \else
  \checkfont{msam10}
  \iffontfound
    \IfFileExists{amssymb.sty}
      {\typeout{^^JFound AMS Symbol fonts on the system, using the
                'amssymb' package.^^J}%
       \usepackage{amssymb}%
         \let\leq=\leqslant
         \let\geq=\geqslant
      }{}
  \fi
\fi

\ifCUPmtlplainloaded \else
  \IfFileExists{amsbsy.sty}
    {\typeout{^^JFound the 'amsbsy' package on the system, using it.^^J}%
     \usepackage{amsbsy}}
    {\providecommand\boldsymbol[1]{\mbox{\boldmath $##1$}}}
\fi

\newcommand\solidrule[1][1cm]{\rule[0.5ex]{#1}{.4pt}}
\newcommand\dashedrule{\mbox{%
  \solidrule[1mm]\hspace{1mm}\solidrule[1mm]\hspace{1mm}\solidrule[1mm]}}
  \newcommand\dottedrule{\mbox{%
  $\cdot$\hspace{1mm}$\cdot$\hspace{1mm}$\cdot$}}
    \newcommand\dasheddottedrule{\mbox{%
  \solidrule[1mm]\hspace{1mm}$\cdot$\hspace{1mm}\solidrule[1mm]}}

\newsavebox{\astrutbox}
\sbox{\astrutbox}{\rule[-5pt]{0pt}{20pt}}

  \newcommand{\rmd}{{\rm d}}
\newcommand{\bx}{{ \boldsymbol{x} }}
\newcommand{\bj}{{ \boldsymbol{j} }}

\newcommand{\by}{{ \boldsymbol{y} }}
\newcommand{\bz}{{ \boldsymbol{z} }}

\newcommand{\bW}{{\wt{\mathbf{W}}}}
\newcommand{\bu}{{ \boldsymbol{u}}}

\newcommand{\bxi}{{\mbox{\boldmath $\xi$}}}

\newcommand{\bsigma}{{\mbox{\boldmath $\sigma$}}}
\newcommand{\hd}{\hat{\rmd}}
\newcommand{\tbxi}{\tilde{\bxi}}

\newcommand{\tell}{\tilde{\ell}}

\newcommand{\bE}{{\mathbb{E}}}
\newcommand{\var}{{\rm Var}}
\newcommand{\wt}{\widetilde}

\newcommand{\tzeta}{{\tilde{\zeta }}}
\newcommand{\teta}{{\tilde{\eta}}}
\newcommand{\tth}{{\tilde{\theta}}}
\newcommand{\txi}{{\tilde{\xi}}}

\newcommand{\ba}{{\bold a}}

\newcommand{\bmu}{{\boldsymbol{\mu}}}

\newcommand{\cD}{{\mathcal{D}}}
\newcommand{\cS}{{\mathcal{S}}}

\newcommand{\red}[1]{\textcolor{red}{#1}}
\newcommand{\blue}[1]{\textcolor{blue}{#1}}

\title[Fluctuation-Dissipation Relation]{A Lagrangian fluctuation-dissipation relation \\
for scalar turbulence, II.  Wall-bounded flows.}

\author[T. D. Drivas and G. L. Eyink]{Theodore D. Drivas$^1$ and Gregory L. Eyink $^{1,2}$ }

\affiliation{$^1$Department of Applied Mathematics \& Statistics, The Johns Hopkins University, Baltimore, MD 21218, USA\\[\affilskip]
$^2$Department of Physics \& Astronomy, The Johns Hopkins University, Baltimore, MD 21218, USA}

\pubyear{2010}
\volume{650}
\pagerange{119--126}
\date{?; revised ?; accepted ?. - To be entered by editorial office}
\begin{document}

\maketitle

\begin{abstract}
{We derive here Lagrangian fluctuation-dissipation relations for advected scalars in wall-bounded flows. The relations equate the dissipation rate for either passive or active scalars to the variance of scalar inputs from the initial values, boundary values, and internal sources, as those are sampled backward in time by stochastic Lagrangian trajectories. New probabilistic concepts are required to represent scalar boundary conditions at the walls: the boundary local-time density at points on the wall where scalar fluxes are imposed and the boundary first hitting-time at points where scalar values are imposed.  These concepts are illustrated both by analytical results for the problem of pure heat conduction and by numerical results from a database of channel-flow turbulence, which also demonstrate the scalar mixing properties of near-wall turbulence. As an application of the fluctuation-dissipation relation, we examine for wall-bounded flows the relation between anomalous scalar dissipation and Lagrangian spontaneous stochasticity, i.e. the persistent non-determinism of Lagrangian particle trajectories in the limit of vanishing viscosity and diffusivity. In the first paper of this series, we showed that spontaneous stochasticity is the only possible mechanism for anomalous dissipation of passive or active scalars, away from walls. Here it is shown that this remains true when there are no scalar fluxes through walls. Simple examples show, on the other hand, that a distinct mechanism of non-vanishing scalar dissipation can be thin scalar boundary layers near the walls. Nevertheless, we prove for general wall-bounded flows that spontaneous stochasticity is another possible mechanism of anomalous scalar dissipation.} 
\end{abstract}

%\begin{keywords}
%Authors should not enter keywords on the manuscript, as these must be chosen by the author during the online %submission process and will then be added during the typesetting process (see http://journals.cambridge.org/data/%\linebreak[3]relatedlink/jfm-\linebreak[3]keywords.pdf for the full list)
%\end{keywords}

%\tableofcontents

\section{Introduction }

{This paper is the second in a series which initiates a new approach to the theory of turbulent scalar dissipation 
based upon a {\it Lagrangian fluctuation-dissipation relation (FDR)}. This relation is derived within a 
representation of the scalar field by stochastic fluid particles, which naturally extends 
Lagrangian methods for ideally advected scalars to realistic problems with both advection and diffusion.
In the first paper of the series (\cite{PaperI}; hereafter, I) the stochastic representation and FDR were obtained for flows 
in finite domains without any boundaries. Using the FDR, we resolved an on-going controversy regarding the phenomenon
of {\it Lagrangian spontaneous stochasticity}, or the persistent stochasticity of Lagrangian trajectories and 
breakdown of uniqueness of particle trajectories for high Reynolds numbers. In ground-breaking work of \cite{Bernardetal98} 
on the \cite{Kraichnan68} model of turbulent advection, the property of Lagrangian spontaneous stochasticity was first discovered
and shown to be a consequence of the classical \cite{Richardson26} theory of turbulent  particle dispersion.   
\cite{Bernardetal98} furthermore showed that anomalous scalar dissipation within the Kraichnan model is 
due to spontaneous stochasticity. The validity of this concept for scalars advected by a true turbulent flow 
has, however, been strongly questioned (e.g. \cite{tsinober2009informal}). A key result of paper I was the demonstration that 
spontaneous stochasticity is the only possible mechanism of anomalous dissipation for both passive and 
active scalars with any advecting velocity field whatsoever, away from walls.}

%\newpage

%\subsection{Wall-bounded domains} 

Wall-bounded flows are{, on the other hand,} ubiquitous both in engineering applications and in nature.  
{Nearly every turbulent flow encountered in our daily experience involves solid walls 
and obstacles that are either stationary or moving, flexible membranes, or free surfaces. Geophysical 
turbulent flows are most commonly constrained by topography, basin boundaries, or surfaces separating 
multiple phases (e.g. air-water). These boundaries can profoundly effect the organization of the turbulence 
and driving from the boundary is very often the origin of the turbulence itself.  Thus, any perspective on turbulent 
scalar dissipation with a claim of generality must be applicable to flows with boundaries. Our aim 
in this work is to extend the approach and results of paper I to the most canonical examples of wall-bounded 
flows. To be specific, we consider a} scalar field $\theta$ (such as temperature, dye concentration, etc.)  
transported by a fluid with velocity $\bu,$ governed by  the advection-diffusion equation in the interior of 
a finite domain $\Omega$:
\begin{align}\label{eq:PS}
\partial_t \theta+\bu\cdot \nabla \theta &= \kappa \Delta \theta + S
\end{align}
with appropriate conditions specified on the boundary $\partial \Omega$.  Here $\kappa>0$ is the molecular diffusivity 
of the scalar and $S(\bx,t)$ is a bulk source field. Typical situations are those 
in which {the} scalar field is held fixed on the boundary (Dirichlet conditions), those in which the scalar flux 
through the wall is fixed (Neumann conditions), or a mixture of these along different parts of the boundary
decomposed as $\partial\Omega=\partial\Omega_D\cup\partial\Omega_N.$ More formally, for a function
$\psi(\bx,t)$ specifying scalar boundary values and a function
$g(\bx,t)$ specifying the fluxes through the wall, one solves 
Eq. \eqref{eq:PS} subject to the conditions 
\begin{alignat}{2}
  \text{Dirichlet:}&\ \ \ \ \ \ \ \ \   &&\theta(\bx,t)= \psi (\bx,t), \quad \bx\in\partial\Omega_D \label{mixed-bc-D} \\
    \text{Neumann:}&\ \ \ \ \ \  \ \ \   -&&{\boldsymbol{\mu}}(\bx,t) \cdot \nabla \theta(\bx,t)= g (\bx,t) \quad \bx\in \partial\Omega_N  
\label{mixed-bc-N} \end{alignat} 
where ${\boldsymbol{\mu}}=\kappa{\bold n}$\footnote{All of our considerations in this paper apply with straightforward modifications
to the more general situation where the scalar flux is given by an anisotropic Fick's law as $\bj(\bx,t)=-{\bold K}(\bx,t)\nabla\theta(\bx,t)$ 
for a positive-definite, symmetric diffusivity tensor ${\bold K}.$ In that case,  the component of the flux normal to the boundary is $j_n={\boldsymbol n}\cdot\bj=-{\boldsymbol{\mu}}\cdot \nabla \theta$ with the so-called ``co-normal vector'' ${\boldsymbol{\mu}}={\bold K}{\bold n}$ satisfying ${\boldsymbol{\mu}} \cdot {\boldsymbol{n}}={\bold n}^\top{\bold K}{\bold n}>0.$}
with $\boldsymbol{n}(\bx)$ being the unit inward normal to the boundary $\partial \Omega$.  Additionally, appropriate boundary conditions
must be specified for the velocity field at the wall.  Throughout this paper, we consider the most common stick (or no-slip) boundary 
conditions:
\begin{equation}
\bu(\bx,t)= {\bf 0}, \quad \bx\in \partial \Omega.
\end{equation}
We shall further discuss in this paper advection only by an incompressible fluid satisfying 
\begin{equation} \nabla\cdot \bu=0, \label{div-free} \end{equation}   
{since only then does fluid advection conserve scalar integrals of the form $\int d^d x \ f(\theta(\bx,t)).$
However, we place no other general constraints on the velocity $\bu,$ which may obey, for example,
an equation of motion involving $\theta$ itself. Thus, our considerations apply not only to passive but also to active scalar fields.}

{For all wall-bounded flows of the above classes we derive Lagrangian fluctuation-dissipation relations generalizing those in paper I 
and we exploit them to precisely characterize for wall-bounded flows the connection between Lagrangian spontaneous 
stochasticity and anomalous scalar dissipation. 
{To avoid confusion, we note that our Lagrangian FDRs have no obvious relation to the fluctuation-dissipation 
theorems traditional in non-equilibrium statistical physics (see discussion in Paper I, Section 5).}
To our knowledge, both these FDR's and the results 
derived for wall-bounded flows have never been discussed previously, as all related works 
(e.g. \cite{Sawfordetal16}) discuss only flows without boundaries. These 
FDR's are obtained, however, within known stochastic representations} for solutions of the scalar advection equation,  
with boundary conditions either of Dirichlet, Neumann, or mixed form (e.g. see \cite{soner2007stochastic}). 
{Our discussion of the stochastic representations themselves involves only very modest originality 
and involves mainly a convenient compilation of existing results from multiple sources. The key concepts 
from probability theory to derive these representations are the {\it boundary local-time density} and the 
{\it boundary hitting-time}, which are carefully described and illustrated with numerical results from a turbulent 
channel-flow simulation. The new FDR's derived within 
these representations} express an exact balance between the time-averaged scalar dissipation and the input 
of stochastic scalar variance from the initial data, interior scalar sources, and boundary conditions as these are sampled 
by stochastic Lagrangian trajectories backward in time. 

With no-flux (zero Neumann) boundary conditions for the scalar, 
we obtain results identical to those of \cite{Bernardetal98} for domains without boundary: namely, spontaneous 
stochasticity is both necessary and sufficient for anomalous dissipation of a passive scalar, and necessary for anomalous 
dissipation of an active scalar.  However, for general imposed fluxes (Neumann conditions), imposed scalar values 
at the wall (Dirichlet conditions) or mixed Dirichlet/Neumann conditions, 
the necessity statement is not generally true. Simple examples, such as pure thermal conduction with imposed heat 
fluxes at the walls, show that thin scalar boundary layers can in such flows provide an entirely distinct 
mechanism of non-vanishing scalar dissipation with a vanishing molecular diffusivity. Nevertheless, we obtain 
from our FDR's a lower bound on the passive scalar dissipation rate, which implies that spontaneous stochasticity is 
sufficient for anomalous dissipation. {With some additional, physically plausible but unproven assumptions, we 
can extend this sufficiency statement also to active scalars. It thus remains true for all wall-bounded flows that 
spontaneous stochasticity is a viable source of anomalous scalar dissipation. This is the main theoretical result
of the present paper. The potential of our FDR's is not, however, exhausted by this single result and our paper 
sets up the framework for more general applications. In the third paper of this series (\cite{PaperIII}; hereafter, III) we apply our 
FDR to turbulent Rayleigh-B\'enard convection and obtain a novel Lagrangian formulation
of the Nusselt-Rayleigh scaling law.} 

{
The detailed contents of the present paper are as follows: in section \ref{Neumann}, we derive the FDR 
in the case of wall-bounded flows with imposed scalar fluxes (\S\ref{StochRep-FDR-Nbc}) and we relate spontaneous 
stochasticity and anomalous dissipation (\S\ref{sec:SS-AnomDiss-Nbc}). The next section \ref{Dirichlet} discusses flows 
with imposed scalar values at the wall and mixed conditions, deriving first an FDR inequality (\S\ref{StochRep-FDR-Dbc}), 
using it to relate spontaneous stochasticity to anomalous dissipation (\S\ref{sec:SS-FDR-Dbc}), then deriving an FDR equality 
(\S\ref{FDR-equal-Dbc}) and lastly discussing mixed boundary conditions (\S\ref{mixed}). The numerical results 
for turbulent channel-flow (\S\ref{DNS-Nbc},\S\ref{DNS-Dbc}) serve not only to illustrate key probabilistic concepts 
for a fluid-mechanics audience but also investigate the joint effects of turbulence and walls on scalar mixing rates, 
which are important in our subsequent study of Rayleigh-B\'enard convection (III).  In the summary 
and discussion section \ref{sec:summary} we discuss some future challenges. 
Three appendices give further details, including analytical results for a 
%\subsection{Proofs for Section \ref{BGK}}\ref{AppBGK}
%\subsection{Proofs for Section \ref{sec:SS-FDR}}\ref{proofsSec2}
%\subsection{Proofs for Section \ref{sec:SS-FDR-Dbc}}\ref{App:A3}
test case of pure diffusion used to assess the effects of fluid advection (Appendix \ref{conduction} and \ref{conductionFixedT}), 
%\subsection{Zero diffusion limit at finite time}\ref{limtemp:finitetime}
%\subsection{Infinite-time limit for fixed diffusivity}\ref{Sol:longtimefixedkappa}
%\subsection{Boundary Local Time Density of Reflected Brownian on the Half-Line}\ref{reflBr},
rigorous details of the new proofs relating spontaneous stochasticity and anomalous dissipation (Appendix \ref{App:A3}),
and explanation of the numerical methods employed for our turbulent channel-flow studies (Appendix \ref{numerics}). 
As in Paper I,  footnotes in the text provide important details for specialists which can probably be ignored by 
a general reader in a first pass through the paper.}

 \section{Imposed Scalar Fluxes at the Wall} \label{Neumann}

We consider in this section a scalar $\theta(\bx,t)$ with pure Neumann boundary conditions ($\partial\Omega_N=\partial\Omega$),
satisfying: 
\begin{alignat}{2}\label{NBC passive scalar}
\partial_t \theta+\bu\cdot \nabla \theta &= \kappa \Delta \theta+ S  \ \ \ \ \ \ \ \ \ &&\bx \in  \Omega,\\
-{\boldsymbol{\mu}}(\bx,t) \cdot \nabla \theta(\bx,t)&= g (\bx,t)  \ \ \ \ \ \ &&\bx \in \partial \Omega. \nonumber
  \end{alignat}
for given initial data $\theta_0,$ {source field $S$}, flux function $g,$ and co-normal vector field ${\boldsymbol{\mu}}.$

\subsection{Stochastic Representation and Fluctuation-Dissipation Relation}\label{StochRep-FDR-Nbc}

The stochastic representation of solutions of pure Neumann problems has been discussed in many earlier publications, such as 
\cite{freidlin1985functional,burdzy2004heat,soner2007stochastic} on fundamental mathematical theory and 
and \cite{mil1996application,keanini2007random,slominski2013weak} for numerical methods. In order to impose Neumann 
conditions, essentially, one averages over stochastic trajectories which reflect off the boundary of the domain, contributing 
to the solution a correction related to their sojourn time on the walls.  We briefly review the subject here. {Our 
presentation is somewhat different than in any of the above references and, in particular, is based on backward stochastic 
integration theory. This framework yields, in our view, the simplest and most intuitive derivations. We 
also specify appropriate physical dimensions for all quantities, whereas the mathematical literature is generally negligent 
about physical dimensions and studies only the special value of the diffusivity $\kappa=1/2,$ in unspecified units. We have been careful
to restore a general value of the diffusivity $\kappa$ in all positions where it appears.}
 
The fundamental notion of the representation is the {\it boundary local time density} $\tell_{t,s}(\bx)$ \citep{stroock1971diffusion,
lions1984stochastic,burdzy2004heat},
%\textcolor{blue}{which represents the amount of time in the interval $[s,t]$ that a stochastic Lagrangian particle located at $\bx\in \Omega$ at time $t$ spends on the boundary $\partial\Omega$. }
which, for a stochastic Lagrangian particle located at $\bx\in \Omega$ at time $t$, is the time within the interval $[s,t]$ which is spent 
near the boundary $\partial\Omega$ per unit distance. Its physical units are thus (time)/(length) or 1/(velocity). 
Note that we consider here the backward-in-time process, {whereas \cite{burdzy2004heat} employ the forward-time 
process and time-reflection to derive an analogous representation}. We thus take the 
boundary local time density to be a non-positive, non-increasing random process which decreases (backward in time) 
only when the stochastic particle is on the boundary. {This stochastic process} is defined via the ``Skorohod problem'', in conjunction 
with the (backward) stochastic flow 
$\tbxi_{t,s}^{\nu,\kappa}(\bx)$ with reflecting b.c. which satisfies 
 \begin{align}\label{stochflowreflected}
\hd \tbxi_{t,s}(\bx)&=\bu^\nu( \tbxi_{t,s}(\bx),s)\ \rmd s+\sqrt{2\kappa}\ \hat{\rmd}\bW_s
-{\boldsymbol{\mu}}(\tbxi_{t,s}(\bx),s) \  \hat{\rmd} \tell_{t,s}(\bx)
\end{align}
with  $\tbxi_{t,t}(\bx)=\bx$ as usual. The boundary local time is then given by the formula 
\begin{equation} \tell_{t,s}(\bx)= \int_t^s dr\ \delta({\rm dist}(\tbxi_{t,r}(\bx),\partial\Omega))\equiv
\lim_{\varepsilon\rightarrow 0}\frac{1}{\varepsilon} \int_t^s dr\ 
\chi_{\partial\Omega_\varepsilon}(\tbxi_{t,r}(\bx)), \qquad s<t, \label{loctimden-form} \end{equation} 
in terms of an ``$\varepsilon$-thickened boundary'' 
\begin{equation} 
\partial\Omega_\varepsilon=\{\bx\in \Omega: {\rm dist}(\bx,\partial\Omega)<\varepsilon \}. 
\end{equation}
See \cite{burdzy2004heat}, Theorem 2.6\footnote{Note that the eq.(2.7) of \cite{burdzy2004heat} contains an 
additional factor of $1/2,$ because their local time density is our $\kappa  \tell_{t,s}(\bx)$ and they consider
only the case $\kappa=1/2$.}.  In (\ref{loctimden-form}) we denote by $\chi_A$ the characteristic function of a set defined as 
\begin{equation}  \chi_A(\bx)=\left\{\begin{array}{ll}
                                  1 & x\in A\cr
                                  0 & x\notin A
                                  \end{array} \right.
   \end{equation}
 \cite{lions1984stochastic}  have proved existence and uniqueness of {stochastic processes as} strong solutions to this ``Skorohod problem''
 with Lipschitz velocity fields $\bu$ and sufficient regular normal vectors ${\bold n}$ at a smooth boundary.
Of some interest for the consideration of non-smooth velocity fields that might appear in the limit 
$\nu\rightarrow 0,$  \cite{stroock1971diffusion} obtain {stochastic process} solutions when $\bu$ is merely bounded, measurable 
and establish uniqueness in law, {i.e. uniqueness of probability distributions}.  
The spatial regularity in $\bx$ of the processes $(\tbxi_{t,s}(\bx),\tell_{t,s}(\bx))$ 
defined jointly as above is the subject of recent works on stochastic flows with reflecting b.c
\citep{pilipenko2005properties,pilipenko2014introduction,burdzy2009differentiability}. Note that 
the stochastic particles described by $\tbxi_{t,s}(\bx)$ are reflected in the direction of $\boldsymbol{\mu}$ 
when they hit the wall, thus staying forever within $\Omega,$ and the flow preserves the volume within the domain
when $\nabla\cdot\bu^\nu=0,$ as assumed here. It follows formally from (\refeq{loctimden-form}) that the boundary 
local-time density can be written as 
\begin{equation}
\tell_{t,s}(\bx)=\int_t^s dr  \int_{\partial\Omega}  dH^{d-1}(\bz) \delta^d(\bz-\tbxi_{t,r}(\bx)), \label{loctime-formal} 
\end{equation} 
where $H^{d-1}$ is the $(d-1)$-dimensional Hausdorff measure (surface area) over the smooth boundary $\partial\Omega$ 
of the domain. We have not found this intuitive formula anywhere in the probability theory literature, but it should be 
possible to prove rigorously with suitable smoothness assumptions on the boundary.

By means of the above notions one can obtain a stochastic representation for solutions to the 
initial-value problem of the system (\ref{NBC passive scalar}). The backward 
It$\bar{{\rm o}}$ formula  \citep{Friedman06,Kunita97} applied to $\theta(\tbxi_{t,s}(\bx),s)$ gives 
\begin{eqnarray}
\hd \theta(\tbxi_{t,s}(\bx),s) &= &\left[(\partial_s\theta +\bu \cdot \nabla \theta-\kappa \Delta \theta) \rmd s -(\nabla \theta \cdot {\boldsymbol{\mu}})  \hd \tell_{t,s}+  \sqrt{2\kappa} \ \hd \bW_s\cdot  \nabla \theta \right]_{(\tbxi_{t,s}(\bx),s) }\cr
&=& S{(\tbxi_{t,s}(\bx),s) }  \rmd s +g{(\tbxi_{t,s}(\bx),s) }  \hd \tell_{t,s}+  \sqrt{2\kappa} \ \hd \bW_s\cdot  \nabla \theta{(\tbxi_{t,s}(\bx),s) } 
\label{back-Ito-refl} \end{eqnarray}
where in the second line we have used the fact that $\theta(\bx,s)$ solves the boundary-value problem (\ref{NBC passive scalar}).
We find upon integration from 0 to $t$,
\begin{eqnarray}\label{back-int-Nbc}
   \theta(\bx,t)&=&  \theta_0(\tbxi_{t,0}(\bx)) + \int_{0}^t\rmd s\ S{(\tbxi_{t,s}(\bx),s) } +\int_{0}^t g{(\tbxi_{t,s}(\bx),s) } \ \hd \tell_{t,s}\cr
  && \ \ \  + \sqrt{2\kappa} \int_{0}^t \hd \bW_s\cdot   \nabla \theta{(\tbxi_{t,s}(\bx),s) } 
\end{eqnarray}
and thus expectation over the Brownian motion yields the representation 
\begin{equation} \label{NeumannStochRep}
\theta(\bx,t)= {\mathbb E}\left[ \theta_0(\tbxi_{t,0}(\bx)) + \int_{0}^t\rmd s\ S{(\tbxi_{t,s}(\bx),s) } 
+\int_{0}^t g{(\tbxi_{t,s}(\bx),s) } \ \hd \tell_{t,s}\right]. \end{equation}
{By means of (\ref{loctime-formal}), this formula can be written equivalently in terms of the transition probability density 
function $p^{\nu,\kappa}(\by,s|\bx,t)={\mathbb E}\left[\delta^d\left(\tbxi_{t,s}^{\nu,\kappa}(\bx)-\by\right)\right]$ for the reflected 
particle to be at position $\by$ at time $s<t,$ given that it was at position $\bx$ at time $t$:} 

\newpage

\begin{eqnarray}
\theta(\bx,t) &=& \int_\Omega d^d\bx_0\ \theta_0(\bx_0) \ p(\bx_0,0|\bx,t) + \int_0^t ds  \int_\Omega  d^d\by\ S(\by,s) \ p(\by,s|\bx,t) \cr
&& \hspace{40pt} +  \int_0^t ds  \int_{\partial\Omega}  dH^{d-1}(\bz)\ g(\bz,s) \ p(\bz,s|\bx,t). 
\end{eqnarray}
Clearly, the scalar value $\theta(\bx,t)$ is an average of the randomly sampled initial data and the 
scalar inputs from the internal source and the scalar flux through the boundary along a random history backward in time. 
  
Just as for domains without walls {in paper I}, we introduce a stochastic scalar field which represents the contribution for 
a specific stochastic Lagrangian trajectory
\begin{equation}\label{tth-Nbc}
\tth(\bx,t)\equiv \theta_0(\tbxi_{t,0}(\bx))
+ \int_{0}^t \rmd s\ S{(\tbxi_{t,s}(\bx),s) } + \int_{0}^t g{(\tbxi_{t,s}(\bx),s)\hd \tell_{t,s}}, 
\end{equation}
that now includes the boundary-flux term, so that $\theta(\bx,t)= {\mathbb E}\left[ \tth(\bx,t)\right]$\footnote{{For all $s<t$
the quantity $\tilde{\theta}(\bx,t;s)=\theta(\tbxi_{t,s}(\bx),s)+ \int_{s}^t S{(\tbxi_{t,r}(\bx),r) } \ \rmd r 
+\int_{s}^t g{(\tbxi_{t,r}(\bx),r) } \ \hd \tell_{t,r}$ is equal to 
$\theta(\bx,t)-\sqrt{2\kappa} \int_{s}^t \hd \bW_r\cdot  \nabla \theta(\tbxi_{t,r}(\bx),r),$ analogous to (\ref{back-int-Nbc}). 
It is thus a martingale back-} {ward in time, i.e. it is statistically conserved and, on average, equals $\theta(\bx,t)$ for 
all $s<t.$ Clearly,} {$\tth(\bx,t)=\tilde{\theta}(\bx,t;0).$ This backward martingale property shows that the stochastic 
representation employed here is a natural generalization to diffusive advection of the standard deterministic 
Lagrangian representation for non-diffusive, smooth advection.}}. {As in paper I,} {we warn the reader that the 
quantity $\tth(\bx,t)$ is entirely different from the conventional ``turbulent'' scalar fluctuation $\theta'(\bx,t)$ defined
with respect to ensembles of scalar initial conditions, advecting velocities, or random sources.  
An application of the It$\bar{{\rm o}}$-isometry 
(see \cite{oksendal2013stochastic}, section 3.1) exactly as in paper I yields for the scalar variance} 
\begin{equation}\label{loc-FDR-Nbc}
 \frac{1}{2}\var\left[  \tth(\bx,t)\right]= \kappa  \ \bE\int_{0}^t \rmd s\   |\nabla \theta{(\tbxi_{t,s}(\bx),s) }|^2 , 
\end{equation}
which is a local version of our FDR. Integrating over the bounded domain and invoking volume preservation by the 
reflected flow, we obtain: 
 \begin{equation}
\frac{1}{2} \left\langle\var\ \tth(t)\right\rangle_\Omega
= \kappa \int_{0}^t \rmd s \left\langle | \nabla\theta(s) |^2\right\rangle_\Omega  \label{FDR-Nbc}
\end{equation}
This, together with (\ref{tth-Nbc}), is our exact FDR for scalars (either passive or active) with 
general Neumann boundary conditions. Just as for {eq.(2.20)} of Paper I in the case of domains without walls, 
we obtain for the infinite-time limit of the local scalar variance 
\begin{equation}\label{loc-FDR-Nbc-long}
\lim_{t\rightarrow\infty} \frac{1}{2t}\var\left[\tth(\bx,t)\right]= 
\left\langle\kappa|\nabla \theta|^2\right\rangle_{\Omega,\infty} ,  \quad \mbox{ for all } \bx\in \Omega,
\end{equation}
{when the reflected stochastic particle wanders ergodically over the flow domain $\Omega$.
This will generally be true when $\kappa>0,$ and the ergodic average thus coincides with the average 
over the stationary distribution of the particle. This is uniform (Lebesgue) measure because of incompressibility
of the flow.  The long-time limit is thus independent of the space point $\bx$. We shall not make use of 
eq.(\ref{loc-FDR-Nbc-long}) in the present work, but it plays a central role in our analysis of steady-state turbulent 
convection in paper III.} 

%%%%%%%%%%%%%%%%%%%%%%%%%%%%%%%%%
%%%%%%%%%%%%%%%%%%%%%%%%%%%%%%%%%
%%%%%%%%%%%%%%%%%%%%%%%%%%%%%%%%%
%NUMERICAL METHODS \& RESULTS

\subsection{Numerical Results}\label{DNS-Nbc} 

To provide some additional insight into these concepts, we present in Appendix \ref{conduction} an elementary analytical example: 
{pure diffusion on a finite interval with a constant scalar flux imposed at the two ends. To be concrete, we shall 
use the language of heat conduction, and thus refer to temperature field, heat flux, etc.} 
 In addition, we present now some numerical results on stochastic Lagrangian trajectories and boundary local time densities for a proto-typical 
 wall-bounded flow, turbulent channel flow. These numerical results are directly relevant to turbulent thermal transport in a channel 
with imposed heat fluxes at the walls. They also have relevance to the problem of Rayleigh-B\'enard convection
which will be discussed in {paper III}, since channel flow provides the simplest example of a turbulent shear 
flow with a logarithmic law-of-the-wall of the type conjectured to exist at very high Rayleigh numbers in 
turbulent convection \citep{kraichnan1962turbulent}.    For the purpose of our study, we use the $Re_\tau=1000$ channel-flow dataset 
in the Johns Hopkins Turbulence Database (see \cite{graham2016web} and \url{http://turbulence.pha.jhu.edu}). We follow 
the notations in these references and, in particular, the $x$ direction is streamwise,  $y$ is wall-normal, and $z$ is spanwise.
For the details of our numerical methods, see  Appendix \ref{channelAppendix}. 
We plot in Figure \ref{fig3} for three choices of Prandtl number ($Pr=$0.1, 1, 10) a single realization of 
$(\tbxi_{t,s}(\bx),\tell_{t,s}(\bx))$ for a stochastic particle released at a point $\bx$ on the lower wall $y=-h$
at the final time $t_f$ in the database, then evolved backward in time. Using Cartesian coordinates $\tbxi_{t,s}=(\txi_{t,s},\teta_{t,s},\tzeta_{t,s}),$ 
the left panel of Fig.~\ref{fig3} plots the wall-normal particle position $\teta_{t,s}(\bx)$ as 
height above the wall $\Delta y=y+h$ and the right panel plots the local time density $\tell_{t,s}(\bx),$ both as functions of 
shifted time $s-t_f.$ All quantities are expressed in wall units, in terms of the friction velocity $u_\tau,$ the viscous length $\delta_\tau=\nu/u_\tau,$
and the viscous time $\tau_\nu=\nu/u_\tau^2.$ Important features to observe in Fig.~\ref{fig3} are the jumps in the local time density 
at the instants when the particles are incident upon the wall and reflected. These incidences occur predominantly near the 
release time and eventually cease (backward in time) as the particles are transported away from the wall. This escape from the wall occurs 
slower for smaller $\kappa$ or larger $Pr,$ and the local times at the wall are correspondingly larger magnitude for larger $Pr.$

\begin{figure}
\centering    
	\includegraphics[width=.48\linewidth]{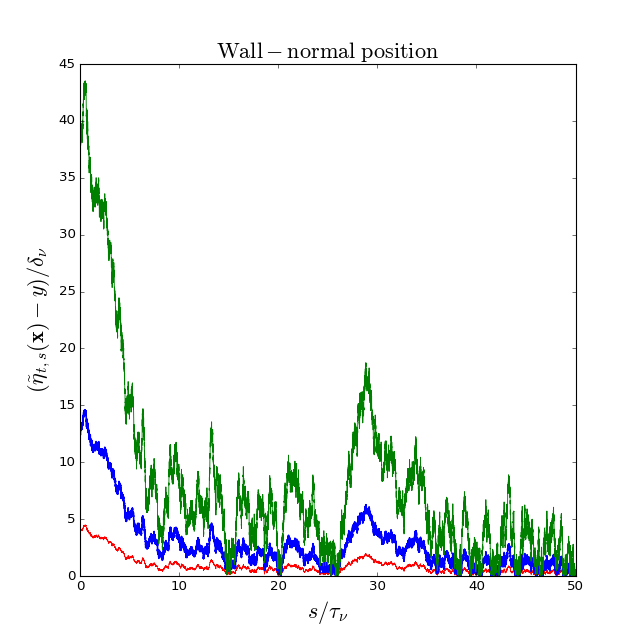}\hspace{-5mm}
	\includegraphics[width=.48\linewidth]{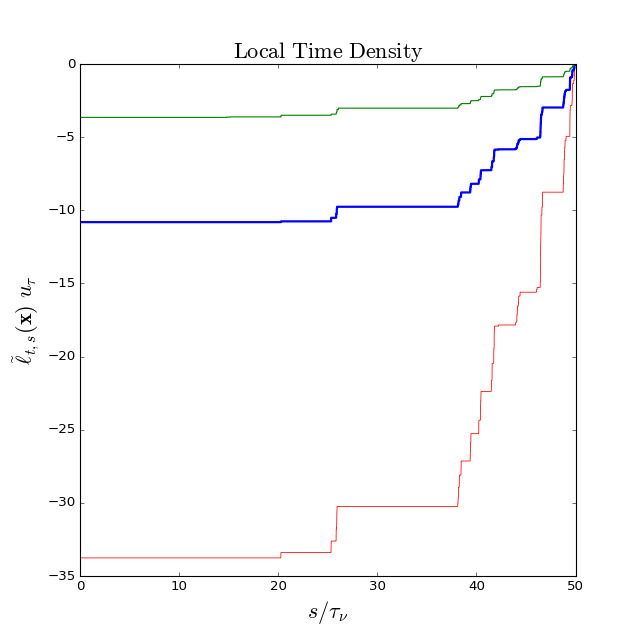}
	\caption{Realizations of wall-normal position (left panel) and local time density process (right panel) for a particle released on the lower 
	wall at $(x,z) = (3543.84\delta_\nu, 1891.56\delta_\nu)$ for Prandtl values $Pr = 0.1$ (\textcolor{ForestGreen}{green}, medium), 
	$1.0$ (\blue{blue}, heavy), and $10$ (\red{red}, light).}
\label{fig3}
\end{figure}

\begin{figure}
\centering    
	\includegraphics[width=.7\linewidth,height=.7\linewidth]{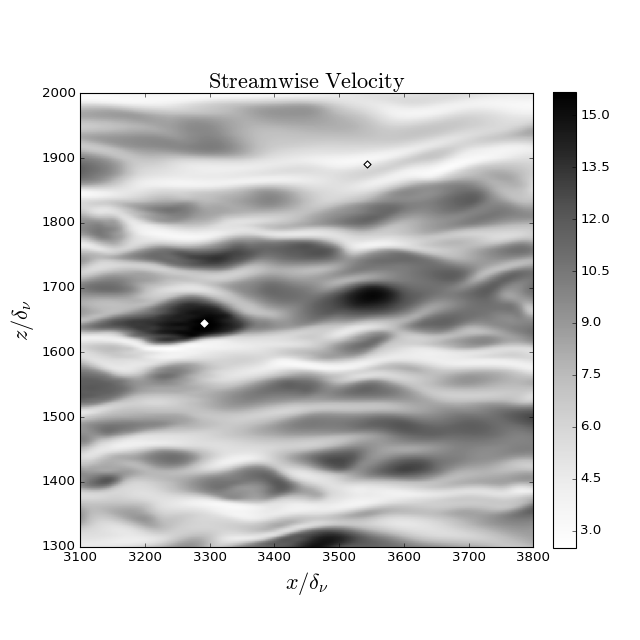}
	\caption{Streamwise velocity at normal distance $\Delta y=10\delta_\nu$ from the lower wall,
	non-dimensionalized by the friction velocity $u_\tau.$  The open diamonds ($\diamond$) mark the $(x,z)$ coordinates,
	 $(3291.73\delta_\nu, 1644.98\delta_\nu)$ and $(3543.84\delta_\nu, 1891.56\delta_\nu),$ of the selected 
	release points at wall-parallel positions of high-speed (dark) and low-speed (light) streaks, respectively.}
\label{fig2}
\end{figure}

We next discuss statistics of the entire ensemble of stochastic particles for two specific release points at the wall, one 
point selected inside a low-speed streak and the other a high-speed streak. {We are especially interested here 
in the role of turbulence in enhancing heat transport away from the wall, compared with pure thermal diffusion.
It has sometimes between questioned whether turbulence close to the wall indeed increases thermal transport, because 
of the restrictions imposed on the vertical motions (e.g. \cite{niemela2003confined}, section 7.2). We thus 
select two points, one in a low-speed streak with mean motion toward the wall (backward in time) and the other 
in a high-speed streak with mean motion away from the wall (backward in time). Comparing these two points helps 
to identify any possible  strictures on turbulent transport imposed by the solid wall.}
The selection is illustrated in Figure \ref{fig2}, 
which plots the streamwise velocity in the buffer layer at distance $\Delta y=10\delta_\tau$ above the bottom channel wall. 
The low-speed streaks associated with ``ejections'' from the wall and high-speed streaks associated with ``sweeps'' 
toward the wall seen there are characteristic of turbulent boundary layers \citep{kline30l967}. The $x$-$z$ coordinates of the two release 
points are indicated by the diamonds ($\diamond$) in Fig.~\ref{fig2} (and note that the point selected in the low-speed streak 
is that featured in the previous Fig.~\ref{fig3}). For each of these two release points and for the three choices of Prandtl number 
$Pr=$0.1, 1, 10, we used $N=1024$ independent solutions of Eqs.(\ref{stochflowreflected}),(\ref{loctimden-form}) 
to calculate the dispersion of the stochastic trajectories and the PDF's of the local-time densities backward in time. 
Error bars in the plots represent s.e.m. for the $N$-sample averages and in addition, 
for the PDF's, the effects of variation in kernel density bandwidth.  

%%%%%%%%%%%%%%%%%%%%%%%%%%%%%%%%%
%%%%%%%%%%%%%%%%%%%%%%%%%%%%%%%%%
%%%%%%%%%%%%%%%%%%%%%%%%%%%%%%%%%

\begin{figure*}
  \begin{subfigure}[b]{0.5\linewidth}
    \centering
    \includegraphics[width=1\columnwidth,  height=.9\columnwidth]{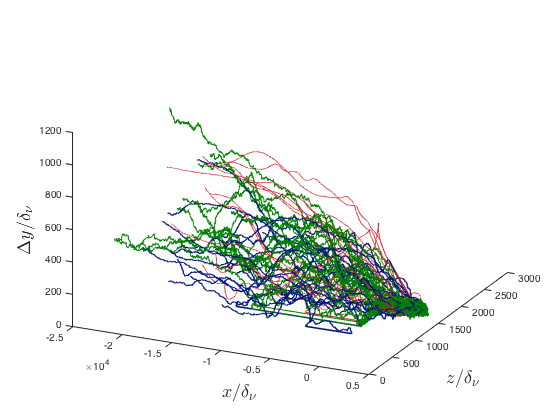} 
      \vspace{-5mm}
   \caption{} 
    \label{fig3:a} 
 \vspace{-1mm}
  \end{subfigure}%% 
  \begin{subfigure}[b]{0.5\linewidth}
    \centering
    \includegraphics[width=1\columnwidth,  height=.9\columnwidth]{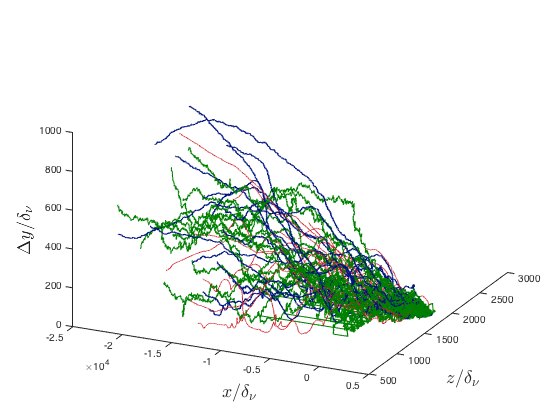} 
      \vspace{-5mm}
   \caption{} 
    \label{fig3:b} 
 \vspace{-1mm}
  \end{subfigure} 

  \begin{subfigure}[b]{0.5\linewidth}
    \centering
    \includegraphics[width=.9\columnwidth,  height=.9\columnwidth]{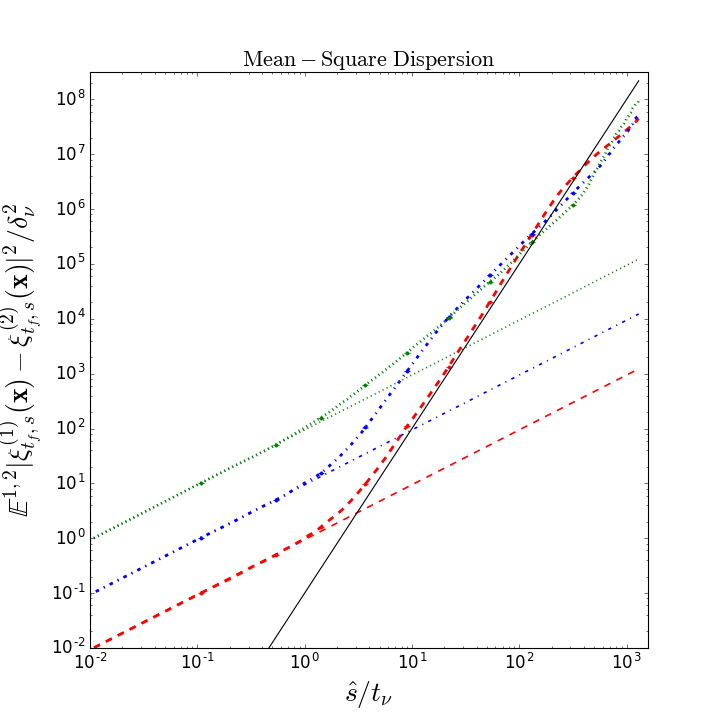} 
      \vspace{-2mm}
   \caption{} 
    \label{fig3:c} 
 \vspace{-1mm}
  \end{subfigure}%% 
  \begin{subfigure}[b]{0.5\linewidth}
    \centering
    \includegraphics[width=.9\columnwidth,  height=.9\columnwidth]{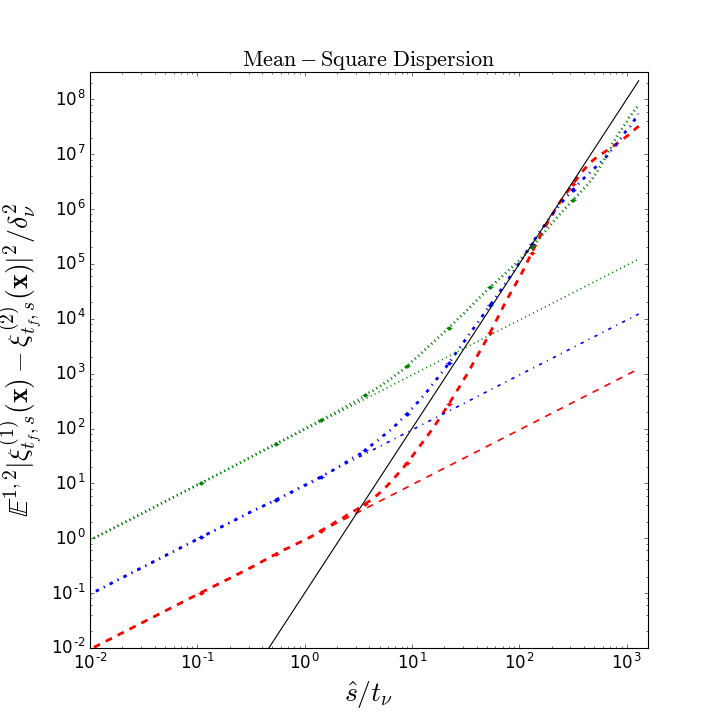} 
      \vspace{-2mm}
   \caption{} 
    \label{fig3:d} 
 \vspace{-1mm}
  \end{subfigure} 

  \begin{subfigure}[b]{0.5\linewidth}
    \centering
    \includegraphics[width=.9\columnwidth,  height=.9\columnwidth]{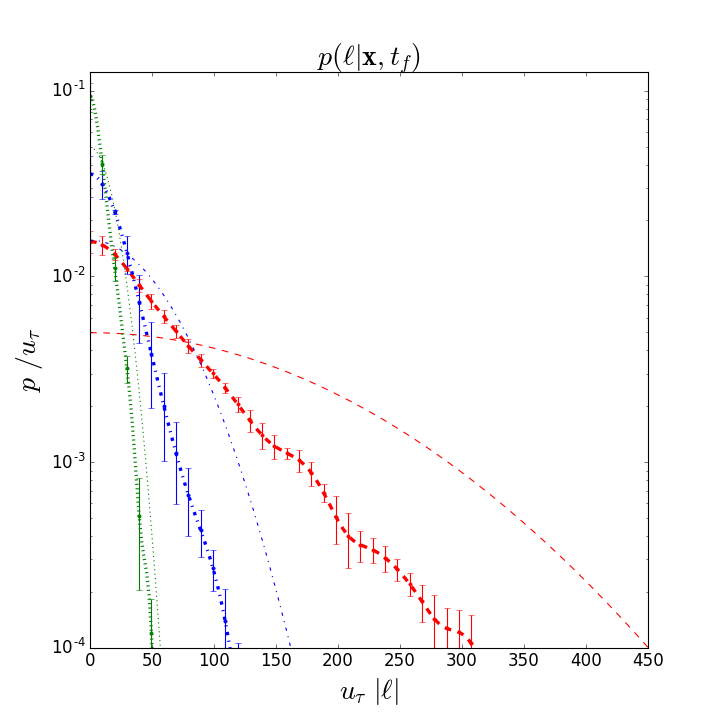} 
      \vspace{-2mm}
   \caption{} 
    \label{fig3:e} 
        \vspace{-1mm}
  \end{subfigure}%% 
  \begin{subfigure}[b]{0.5\linewidth}
    \centering
    \includegraphics[width=.9\columnwidth,  height=.9\columnwidth]{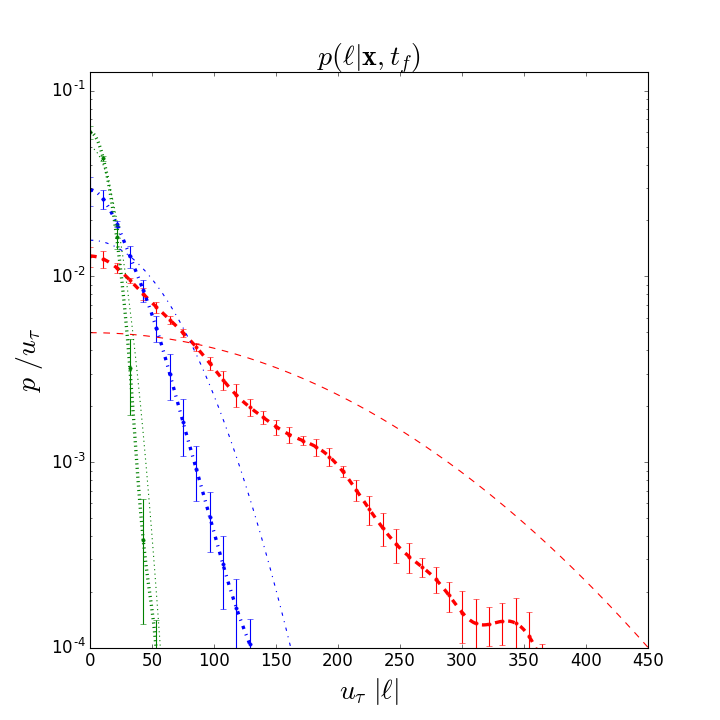} 
   \vspace{-2mm}
   \caption{} 
    \label{fig3:f} 
  \end{subfigure} 
	\caption{\emph{Left panels} are for release at bottom wall near a high-speed streak, \emph{right panels} for release near 
	a low-speed streak.  See markers on Figure \ref{fig2}.  Top panels (a),(b) show 30 representative reflected stochastic 
	trajectories for $Pr = 0.1$ (\textcolor{ForestGreen}{green}, medium), 
	$1.0$ (\blue{blue}, heavy), and $10$ (\red{red}, light).  Middle panels 
	(c),(d) plot particle dispersion (heavy) and short-time results $4\kappa(3-\frac{2}{\pi})\hat{s}$ (light) for each $Pr$ with $Pr = 0.1$ (\textcolor{ForestGreen}{green}, dot, \dottedrule), $1.0$ (\textcolor{blue}{blue}, dash-dot, \dasheddottedrule) and $10$  (\textcolor{red}{red}, dash, \dashedrule) and a plot (black,solid, \solidrule[4mm]) of $g (\nu/\tau_\nu^2)\hat{s}^3$ with $g=0.1$. 
	 The bottom panels (e),(f) plot PDF's of 
	(absolute values of) boundary local times for channel flow (heavy) and pure diffusion (light) for the three $Pr$-values with the same line-styles as (c),(d).}
\label{fig:channel}
\end{figure*}

In Figure \ref{fig:channel}, the top panels show 30 representative particle trajectories for the two release points (left within a high-speed streak, 
right in a low-speed) and for each of the three Prandtl numbers.  By eye, the ensembles of trajectories for the three different Prandtl numbers appear 
quite similar.  In the middle panels of Fig. \ref{fig:channel}, we plot the mean-squared dispersion of the stochastic particles for the two release 
points and the three Prandtl numbers. There is an initial period (backward in time) where the dispersion grows 
with $\hat{s} = t_f-s$ as $4\kappa(3-2/\pi) \hat{s},$ which is the analytical result for a 3D Brownian motion reflected from a plane\footnote{
If $\tilde{W}(t)$ is standard 1D Brownian motion with diffusivity $\kappa$, then $|\tilde{W}(t)|$ is the Brownian motion reflected 
at 0 and it is elementary to show that $\bE^{1,2}\big[\{|\tilde{W}^{(1)}(t)|-|\tilde{W}^{(2)}(t)|\}^2\big]=4\kappa(1-2/\pi)t.$}.
There is then a crossover to a limited regime of super-ballistic separation 
that is close to $\hat{s}^3$-growth. At still larger $\hat{s},$ the growth clearly slows but remains super-ballistic.
{Recall from Paper I that Richardson dispersion with a cubic growth of dispersion is the physical 
mechanism of spontaneous stochasticity in homogeneous, isotropic turbulence.} It would be quite 
surprising to observe Richardson dispersion in this channel-flow dataset, since the energy spectra published 
in \cite{graham2016web} and \url{http://turbulence.pha.jhu.edu} show that the Reynolds number is still too low to support a 
Kolmogorov-type inertial-range. We believe that the cubic power-law growth observed arises instead from a simple combination 
of stochastic diffusion and mean shear, a molecular version of the turbulent shear dispersion discussed by \cite{corrsin1953remarks}.  
A toy model which illustrates this mechanism is a 2D shear flow with $\bu(\bx)=(Sy,0),$ where 
\begin{equation}  d\txi_t= S \teta_t + \sqrt{2\kappa}\ d\tilde{W}_x(t),\quad  d\teta_t = \sqrt{2\kappa}\ d\tilde{W}_y(t).
\label{shear-disp} 
\end{equation}
Integration gives $\txi_t =\sqrt{2\kappa}( S\int_0^t \tilde{W}_y(s) ds + \tilde{W}_x(t))$ and $\teta_t = 
\sqrt{2\kappa}\ \tilde{W}_y(t)$ when $(\txi_0,\teta_0)=(0,0),$ 
and the shear contributes a $t^3$-term to the $x$-component of the dispersion:
\begin{equation} \mathbb{E}^{1,2}|\txi^{(1)}_{t}-\txi^{(2)}_{t}|^2 = 4\kappa(t + \frac{1}{3} S^2  t^3). \label{toy-disperse} \end{equation} 
{See \cite{schutz2016} for a very similar discussion of cubic-in-time dispersion in the context of bounded shear flows and 
weakly turbulent Rayleigh-B\'{e}nard convection in 2D.} 
In the toy model (\ref{shear-disp}) the $y$-component of dispersion grows only diffusively. In qualitative agreement with this simple model,
we have found that the cubic power-law growth in the channel-flow particle dispersion indeed arises solely from the streamwise 
component of the particle separation vector. We do not show this here, but the effect can be observed in the particle trajectories plotted in the top
panels of Fig.~\ref{fig:channel}, where the particles are dispersed farthest along the streamwise direction. The $t^3$ dispersion 
in \eqref{toy-disperse} differs notably from Richardson dispersion in that it is proportional to $\kappa$ and it produces no spontaneous stochasticity,
since it vanishes as $\kappa\rightarrow 0.$ In the channel-flow $\hat{s}^3$-regime in Fig.~\ref{fig:channel}c,d
($10^1\lesssim \hat{s}/t_\eta\lesssim 10^2$) there is likewise an observable dependence on the diffusivity.
{These cautionary results are an admonition against interpreting any cubic or super-ballistic growth 
of pair-separation whatsoever as evidence of spontaneous stochasticity. Richardson dispersion produces spontaneous 
stochasticity because of non-smooth relative advection, not simply because of super-ballistic separation.}

In the bottom panels of Figure \ref{fig:channel}, we plot PDF's of the local time density $\tell_{t_f,0}(\bx)$ accumulated 
backward over the entire time interval of the channel-flow database (one flow-through time), for the two release points 
and the three values of Prandtl number.  As could be expected, smaller $\kappa$ (or larger $Pr$) correspond to more 
time at the wall and a PDF supported on larger local time densities.  Comparing the results for the two release points, 
it can be seen that particles starting on the wall inside the high-speed streak tend to spend somewhat less time near 
the wall than those starting near the low-speed streak.  This is easy to understand, if one recalls that low-speed streaks 
correspond to ``ejections'' from the wall and high-speed streaks to ``sweeps'' toward the wall. Backward in time, however, 
the fluid motions are exactly reversed.  Thus, particles starting at the wall inside high-speed streaks are being swept away 
from the wall by the fluid motion backward in time, and those starting inside low-speed streaks are brought toward the wall. 
However, the location of the release point has a quite small effect on the PDF of the local-time density accumulated over 
one flow-through time, and it is plausible that this effect would be reduced by taking $t_f$ even larger. 
%could vanish entirely in the limit $t_f\rightarrow\infty.$ 

Finally, we have 
also plotted in the bottom panels of Fig.~\ref{fig:channel} the analytical results for the local time PDF's of a pure Brownian 
motion with diffusivity $\kappa$ that is reflected from a planar wall (see Appendix \ref{reflBr} and \eqref{loctim-pdf} below).
Compared with these results for Brownian motion, the corresponding channel-flow local-time PDF's exhibit in all 
cases a substantial reduction of time spent at the wall. These results show that turbulent fluctuations enhance transport 
away from the wall, even when the local flow initially carries the particles toward the wall (backward in time). Although we show 
these results only for two release points here, we have observed similar behavior at all other locations on the wall that 
we have examined. This observation has importance for the problem of turbulent convection, where it was
it was argued by \cite{kraichnan1962turbulent} that  ``\dots the effect of the small-scale turbulence that arises locally in 
the shear boundary layer should be to increase heat transport." We shall return to this issue in paper III. 

\subsection{Spontaneous Stochasticity and Anomalous Dissipation}\label{sec:SS-AnomDiss-Nbc}

As an application of our FDR (\ref{FDR-Nbc}) for flows with imposed scalar fluxes at the walls,
we now discuss the equivalence between spontaneous stochasticity and anomalous scalar dissipation in this context. 
As we shall see, the effects of the walls can be profound.  

The simplest situation is the case of vanishing wall-fluxes, $g\equiv 0,$ which corresponds to {\it insulating/adiabatic walls} 
when the scalar is the temperature and to {\it impermeable walls} when the scalar is the concentration of an advected 
substance (e.g. a dye, aerosol, etc.). In this case, the flux contribution proportional to the local time density in 
(\ref{tth-Nbc}) vanishes, and our FDR then becomes simply
 \begin{align}\label{NeumannVarInsol}
\frac{1}{2} \left\langle\var\left[  \theta_0(\tbxi_{t,0})+ \int_{0}^t\rmd s \ S{(\tbxi_{t,s},s) } \right]\right\rangle_\Omega &
=\kappa \int_{0}^t \rmd s \left\langle | \nabla\theta(s) |^2\right\rangle_\Omega. \end{align}
This is formally identical to the relation {(2.9),{(2.12)} of Paper I} for flows in domains without boundaries, 
with the simple stochastic flow replaced by a reflected stochastic flow.  We can therefore repeat {\it verbatim} the arguments 
from {Paper I, section 4}
to conclude that, for walls that do not support fluxes, spontaneous stochasticity is equivalent to anomalous dissipation 
for passive scalars and for active scalars spontaneous stochasticity is (at least) necessary for anomalous 
scalar dissipation. {For the sake of completeness, we briefly review the main ideas of this argument here.
Assuming that $S\equiv 0$ for simplicity, the lefthand side 
of the FDR (\ref{NeumannVarInsol}) becomes 
\begin{eqnarray}\label{Var2}
&&  {\rm Var}\left[\theta_0(\tbxi_{t,0}(\bx))\right]
   =  \int d^dx_0\int d^dx_0' \ \theta_0(\bx_0) \theta_0(\bx_0') \cr
&&\hspace{35pt}    \times \Big[ p^{\nu,\kappa}_2(\bx_0,0;\bx_0',0|\bx,t)-p^{\nu,\kappa}(\bx_0,0|\bx,t)p^{\nu,\kappa}(\bx_0',0|\bx,t)\Big],
\end{eqnarray}
where we have introduced the 2-time (backward-in-time) transition density
\begin{equation}
p^{\nu,\kappa}_2(\by,s;\by',s'|\bx,t)=
\bE\left[\delta^d(\by-\tbxi_{t,s}^{\nu,\kappa}(\bx))\delta^d(\by'-\tbxi_{t,s'}^{\nu,\kappa}(\bx))\right], \quad s,s'<t.
\label{pdf2-def}\end{equation}  
{It can be shown using Young measure techniques that, at least along suitably chosen 
subsequences, the transition probabilities approach 
limiting values $p^*(\by,s;\by',s|\bx,t)$, $p^*(\by,s|\bx,t)$ as $\nu,\kappa\to 0$ whenever the flow domain is compact 
(Paper I, Appendix A.1). 
If in this limit the Lagrangian particle positions $\bxi_{t,s}^*$ are deterministic, then the 2-time transition probability factorizes:  
\begin{equation}
p_2^*(\by,s;\by',s'|\bx,t)=\delta^d(\by-\bxi_{t,s}^*(\bx))\delta^d(\by'-\bxi_{t,s'}^*(\bx))=p^*(\by,s|\bx,t)p^*(\by',s'|\bx,t). 
\end{equation}}
$\!\!\!\!$ Thus, non-factorization in the limit $\nu,\kappa\rightarrow 0$ is the signature of spontaneous stochasticity,  
i.e. of the limiting particle positions $\bxi_{t,s}^*$ remaining random as $\nu,\kappa\rightarrow 0$. 
Note that the variance (\ref{NeumannVarInsol})  can only be non-vanishing 
in the limit if factorization fails.  Consequently, anomalous dissipation requires spontaneous stochasticity. This is true 
both for passive and for active scalars.  In the other direction, if there is spontaneous stochasticity on a positive-measure 
set of $\bx\in \Omega$, one can choose (at least for passive scalars) a suitable $\theta_0$ which will make the variance (\ref{NeumannVarInsol}) positive.  This implies a positive lower bound to the cumulative, volume-integrated scalar 
dissipation through the FDR (\ref{NeumannVarInsol}). Thus spontaneous stochasticity and anomalous scalar dissipation 
are seen to be equivalent. See Paper I, Appendix A.2 for detailed mathematical proofs of these claims.}

This result is relevant for many physical situations, such as a dye injected into a 
turbulent Taylor-Couette flow. The commonplace example of cream stirred into coffee is also essentially of 
this type, since the cup walls and surface of the stirrer are impermeable boundaries and there is also no transport of cream 
across the free fluid surface. The extension of our FDR (\ref{NeumannVarInsol}) to problems with moving walls and 
free-fluid surfaces should be relatively straightforward (and, indeed, the main result of the paper of \cite{burdzy2004heat} 
was the construction of reflected Brownian processes for domains enclosed by moving boundaries). In all of these situations, 
any evidence for anomalous scalar dissipation is also evidence for (requires) spontaneous stochasticity. 

The situations with non-vanishing fluxes through the walls are, however, essentially different. From a mathematical 
point of view, the problem is ``loss of compactness.'' While the trajectories of the reflected diffusion process 
 satisfy uniformly in $\nu,\kappa$ the condition that $\tbxi^{\nu,\kappa}_{t,s}(\bx)\in\bar{\Omega},$ 
 a closed, bounded domain, the local time densities $\tell_{t,s}^{\nu,\kappa}(\bx)$ may become unboundedly 
 large as $\nu,\kappa\rightarrow 0.$ This creates a fundamental difficulty for arguments of the type employed previously, 
 where limits as $\nu,\kappa\rightarrow 0$ were guaranteed to exist (along subsequences). To understand better
the problem, consider how we might try to adapt those arguments to the present context. We can   
rewrite our FDR (\ref{FDR-Nbc}),(\ref{tth-Nbc}) as 
\begin{eqnarray}
&& \kappa \int_{0}^t \rmd s \left\langle | \nabla\theta(s) |^2\right\rangle_\Omega =
\frac{1}{2} \left\langle\var\left[  \tth^{\nu,\kappa}(t)\right] \right\rangle_\Omega \cr
&& \,\,\,\,
= \frac{1}{2} \left\langle\int d\psi \int d\psi' \ \psi\psi' \ \Big[p_2^{\nu,\kappa}(\psi,\psi'|\cdot,t)
-p^{\nu,\kappa}(\psi|\cdot,t)p^{\nu,\kappa}(\psi|\cdot,t)\Big]\right\rangle_\Omega, 
\end{eqnarray} 
where
\begin{equation}  p^{\nu,\kappa}(\psi|\bx,t)=\bE\Big[\delta(\psi-\tth^{\nu,\kappa}(\bx,t))\Big] 
\end{equation}        
and 
\begin{equation}  p_2^{\nu,\kappa}(\psi,\psi'|\bx,t)=\bE\Big[\delta(\psi-\tth^{\nu,\kappa}(\bx,t))
       \delta(\psi'-\tth^{\nu,\kappa}(\bx,t)\Big] =\delta(\psi-\psi')p^{\nu,\kappa}(\psi|\bx,t), 
       \end{equation}
with $\tth^{\nu,\kappa}(\bx,t)$ given by (\ref{tth-Nbc}). If weak limits 
$p^*(\psi|\bx,t)=w\mbox{-}\lim_{\nu,\kappa\rightarrow 0} p^{\nu,\kappa}(\psi|\bx,t)$ existed, then arguments of exactly 
the type given before would show that anomalous dissipation requires non-factorization of 
$p_2^*$ into $p^*\cdot p^*$ and, hence, spontaneous stochasticity. The reverse argument that 
spontaneous stochasticity implies anomalous dissipation for passive scalars would also be essentially the same.  
            
Simple examples show, however, that weak limits $p^*(\psi|\bx,t)=w$-$\lim_{\nu,\kappa\rightarrow 0} p^{\nu,\kappa}(\psi|\bx,t)$
may not exist when wall-fluxes are not vanishing. Consider the case where $\theta_0=S\equiv 0$ and 
the flux into the domain is a space-time constant $g(\bx,s)=J>0.$ In that case 
\begin{equation}  \tth(\bx,t) =   \int_{0}^t g{(\tbxi_{t,s}(\bx),s)   \hd \tell_{t,s}}(\bx) = -J \tell_{t,0}(\bx)\geq 0,
\end{equation}
and the stochastic scalar field for one realization of the Brownian process is proportional by a constant 
to the local time density itself. Consider furthermore the simple case of pure heat-conduction, where  
the advecting velocity field also vanishes, $\bu^\nu\equiv{\bf 0}.$ In this case, the distribution of $\tell_{t,0}(\bx)$
is known analytically in many cases. A very simple example which serves to make our point takes the 
domain to be the semi-infinite one-dimensional interval $\Omega=[0,\infty)$ with boundary at 0. In that case, for $x\geq 0$
\begin{equation}
p^\kappa(\psi|x,t) = \frac{1}{J}\sqrt{\frac{\kappa}{\pi t}}\exp\left[-\frac{(x+\kappa\psi/J)^2}{4\kappa t}\right]\eta(\psi)
+ \left[2\Phi_{\kappa,t}(x)-1\right]\delta(\psi) \label{loctim-pdf} \end{equation} 
where $\eta(\psi)$ is the Heaviside step function and 
\begin{equation} \Phi_{\kappa, t}(x) = \frac{1}{\sqrt{4\kappa\pi t}}\int_{-\infty}^x dy\ \exp(-y^2/4\kappa t). \end{equation}
For details see Appendix \ref{reflBr}. It is easy to see mathematically from the above expression 
that weak limits exist for fixed $x>0$ and, indeed, 
\begin{equation}  w\mbox{-}\lim_{\kappa\rightarrow 0} p^\kappa(\psi|x,t) = \delta(\psi), \quad x>0. \end{equation}
This is also intuitively obvious, because a stochastic particle released at $x>0$ never makes it to the boundary at $0$
when $\kappa\rightarrow 0.$ However, {\it the distribution $p^\kappa(\psi|0,t)$ does not converge weakly 
as $\kappa\rightarrow 0$}, but (\ref{loctim-pdf}) implies instead that it  tends to become uniformly spread on the semi-infinite 
interval. This also makes sense, because a particle released at 0 should tend to stay there as $\kappa\rightarrow 0$
and the local time density at 0 will diverge\footnote{In fact, if we consider the 1-point compactification $[0,\infty]$
of the range $[0,\infty)$ of possible scalar values, then $\lim_{\kappa\rightarrow 0}p^\kappa(\psi|0,t)=\delta_\infty(\psi),$
the delta-function at infinity. However, this compactification of the problem does not yield any result on spontaneous stochasticity.}. 

The physical origin of the divergence in this simple example is a {\it scalar boundary layer} near $x=0.$ A direct 
solution of the Neumann problem for the scalar field or integration over the above distribution shows that 
\begin{equation} \theta(x,t) = -J\bE[\tell_{t,0}(x)] \sim J \sqrt{\frac{t}{\kappa}} f\left(\frac{x}{\sqrt{\kappa t}}\right)\end{equation}
for a suitable scaling function $f.$ See Appendix \ref{limtemp:finitetime}. There is thus a scalar boundary layer of thickness
$\sim \sqrt{\kappa t}$ near 0 where the scalar field diverges as $\theta\sim J\sqrt{t/\kappa}$ as $\kappa\rightarrow 0.$ Because 
there is a constant flux $J$ into the domain and diffusive transport into the interior vanishes as $\kappa\rightarrow 0,$
there is a ``pile-up'' of the scalar near $x=0.$ The scalar dissipation field due to this boundary layer is 
\begin{equation}  \varepsilon_\theta(x,t) = \kappa |\partial_x\theta(x,t)|^2 \sim \frac{J^2}{\kappa} \left|f' \left(\frac{x}{\sqrt{\kappa t}}\right) \right|^2, \end{equation}
from which one can infer a total scalar dissipation $\propto J^2 \sqrt{t/\kappa}\rightarrow \infty$ as $\kappa\rightarrow 0.$
This ``dissipative anomaly'' occurs even though there is clearly no spontaneous stochasticity in this simple example with 
$\bu^\nu\equiv 0!$

The above example shows that the ``loss of compactness'' is not a mere technical mathematical difficulty, but instead 
that there may no longer be equivalence of spontaneous stochasticity and non-vanishing dissipation.  Scalar boundary layers 
in wall-bounded flow domains with flux through the walls are a new possible source of scalar dissipation that can be 
non-zero (or even diverging) as $\nu,\kappa\rightarrow 0,$ quite distinct from the spontaneous stochasticity mechanism. 
%Of course, our FDR (\ref{FDR-Nbc}),(\ref{tth-Nbc}) is valid even when there is either no spontaneous stochasticity or no anomalous 
%dissipation. As we shall now discuss, the FDR can give physical information on the mechanisms and asymptotic 
%behaviour of scalar dissipation also in such cases. 
{Another perhaps more elementary way to see the problem posed by wall-fluxes is to exploit formula 
(\ref{loctime-formal}) for the boundary local-time to write 
\begin{eqnarray}\label{Var-g}
&&  {\rm Var}\left[\int_{0}^t g{(\tbxi_{t,s}(\bx),s) } \ \hd \tell_{t,s}(\bx) \right]
   =  \int_{0}^tds\int_{0}^tds' \int dH^{d-1}(z)\int dH^{d-1}(z') \ g(\bz,s) g(\bz',s') \cr
&&\hspace{35pt}    \times \Big[p_2^{\nu,\kappa}(\bz,s;\bz',s'|\bx,t)-p^{\nu,\kappa}(\bz,s|\bx,t)p^{\nu,\kappa}(\bz',s'|\bx,t)\Big],
\end{eqnarray} 
where $p^{\nu,\kappa}_2(\by,s;\by',s'|\bx,t)$  for $s,s'<t$ is the 2-time transition probability density function with respect to 
Lebesgue measure, {defined in eq.\eqref{pdf2-def}}. On the face of it, this appears to be quite similar to the analogous 
formulas for the scalar variance associated to the initial data or internal sources, e.g. eq.\eqref{Var2}.
%e.g. 
%\begin{eqnarray}\label{Var2}
%&&  {\rm Var}\left[\theta_0(\tbxi_{t,0}(\bx))\right]
%   =  \int d^dx_0\int d^dx_0' \ \theta_0(\bx_0) \theta_0(\bx_0') \cr
%&&\hspace{35pt}    \times \Big[ p^{\nu,\kappa}_2(\bx_0,0;\bx_0',0|\bx,t)-p^{\nu,\kappa}(\bx_0,0|\bx,t)p^{\nu,\kappa}(\bx_0',0|\bx,t)%\Big]. 
%\end{eqnarray}
However, in the variance associated to $\theta_0$ or $S,$ the transition probability densities appear only in the combinations 
$d^dy\ d^dy'\ p^{\nu,\kappa}_2(\by,s;\by',s'|\bx,t)$ and $d^dy\ p^{\nu,\kappa}(\by,s|\bx,t),$ and thus only involve the 
{\it probability measures} themselves and not their densities. Compactness arguments suffice to show that limits of the probability 
measures and associated integrals exist (along subsequences) as $\nu,\kappa\rightarrow 0.$ On the other hand, the formula 
(\ref{Var-g}) involves the probability density function evaluated on the boundary of the domain. Even if the (weak) limit 
of the particle probability measures have density functions $p^{*}_2(\by,s;\by',s'|\bx,t)$ and $p^{*}(\by,s|\bx,t),$ they are 
undefined at the boundary $\partial \Omega,$ which is a Lebesgue zero-measure set. In general, the 
probability density functions will not have pointwise limits as $\nu,\kappa\rightarrow 0$ and, in the example 
of pure diffusion on the half-line discussed above, they diverge when all three points $\bx,$ $\by$, $\by'$ are located on 
the left boundary at $0$!}   

{On the other hand, we can extend all of our results relating spontaneous stochasticity and anomalous scalar
dissipation to general choices of wall-fluxes, if we make the additional assumption that pointwise limits of densities exist for all 
$\bx,$ $\by,$ $\by'\in \Omega$
\begin{equation}\label{point-lim}
\lim_{\nu,\kappa\rightarrow 0} p^{\nu,\kappa}_2(\by,s;\by',s'|\bx,t)=p^{*}_2(\by,s;\by',s'|\bx,t), \quad
\lim_{\nu,\kappa\rightarrow 0} p^{\nu,\kappa}(\by,s|\bx,t)=p^{*}(\by,s|\bx,t), 
\end{equation} 
so that the limit of the variance in (\ref{Var-g}) exists and is given by the formula
\begin{eqnarray}\label{Var-g-star}
&&  \lim_{\nu,\kappa\rightarrow 0}{\rm Var}\left[\int_{0}^t g{(\tbxi_{t,s}(\bx),s) } \ \hd \tell_{t,s}(\bx) \right]
   =  \int_{0}^tds\int_{0}^tds' \int dH^{d-1}(z)\int dH^{d-1}(z')  \cr
&&\hspace{35pt}    \times g(\bz,s) g(\bz',s')\ \Big[p_2^{*}(\bz,s;\bz',s'|\bx,t)-p^{*}(\bz,s|\bx,t)p^{*}(\bz',s'|\bx,t)\Big].
\end{eqnarray} 
This assumption, of course, rules out the presence of scalar boundary layers of the type discussed above.  
The necessity of spontaneous stochasticity for anomalous scalar dissipation is immediate, because factorization 
$p^{*}_2(\by,s;\by',s'|\bx,t)=p^{*}(\by,s|\bx,t)p^{*}(\by',s'|\bx,t)$ of the limit densities implies that all variances 
tend to zero. If the scalar is passive, then sufficiency holds as well. Consider, for simplicity, the case of vanishing
internal source $S\equiv 0.$ As in Paper I, section 4 and Appendix A, we can make a smooth choice 
of scalar initial data $\theta_0$ so that the volume-average of the limiting variance in (\ref{Var2}) is strictly positive, 
whenever non-factorization holds for a positive measure set of $\bx\in\Omega$. The limit variance 
in (\ref{Var-g-star}) involving the boundary flux $g$ is non-negative when it exists at all. If the covariance 
term involving both $\theta_0$ and $g$ is negative, then by taking $\theta_0\rightarrow -\theta_0 $ the 
covariance can be made positive without changing the sign of either of the two variances involving $\theta_0$
and $g$ separately. In this manner, for any possible choice of wall-fluxes $g,$ an initial condition $\theta_0$
for the passive scalar advection equation always exists so that scalar dissipation is non-vanishing as 
$\nu,\kappa\rightarrow 0.$ The same conclusion holds also for an active scalar, barring a ``conspiracy'' in which 
all Lagrangian particles at time $0$ originating from a.e. point $\bx\in \Omega$ at time $t$ are confined to a single isoscalar 
surface of $\theta_0$ for that $\bx,$ even when varying over all possible $\theta_0.$ It is this unlikely 
coincidence which must be ruled out by a rigorous proof. In this manner we see that all of our previous conclusions
for flows without walls or for zero fluxes at the walls, can be extended to the case of general wall-fluxes $g,$ whenever the 
stringent assumption (\ref{point-lim}) on pointwise limits is valid.} 

\section{Imposed Scalar Values at the Wall and Mixed Conditions} \label{Dirichlet}

In this section we extend our previous results as far as possible to advection-diffusion of scalars in wall-bounded domains 
with the general mixed Dirichlet-Neumann conditions (\ref{mixed-bc-D}),(\ref{mixed-bc-N}). 
 {It turns out that imposing scalar 
values (Dirichlet conditions) on even part of the boundary leads to much more essential difficulties than imposing 
fluxes (Neumann conditions).  In particular, the arguments used earlier for deriving a Lagrangian fluctuation-dissipation 
relation no longer allow us to express the total volume-integrated scalar dissipation in purely Lagrangian terms.
Instead, we can derive two distinct FDR's, one relation giving a lower bound on total scalar dissipation in purely Lagrangian
terms and another relation of mixed Euler-Lagrangian character for the total scalar dissipation. The inequality 
relation allows us to deduce that anomalous dissipation can result from spontaneous stochasticity, but not  that 
spontaneous stochasticity is necessary\footnote{{In fact, we provide an analytical example of  pure 
heat conduction exhibiting a dissipative anomaly due entirely to thin scalar boundary layers 
(Appendix \ref{conductionFixedT}).}}.   
The equality relation is more useful for physical purposes. We thus 
derive both relations.} We first consider the special case of a 
scalar $\theta(\bx,t)$ with pure Dirichlet boundary conditions ($\partial\Omega_D=\partial\Omega$):
\begin{alignat}{2}\label{DBC passive scalar}
\partial_t \theta+\bu\cdot \nabla \theta &= \kappa \Delta \theta+ S  \ \ \ \ \ \ \ \ \ &&\text{ for } \ \ \ \bx \in  \Omega,\\
  \theta(\bx,t)&= \psi (\bx,t)   \ \ \ \ \ \ &&\text{ for } \ \ \ \bx \in \partial \Omega. \nonumber 
  \end{alignat}
for given initial data $\theta_0,$ boundary data $\psi,$ {and source field $S(\bx,t)$.} Of course, pure Dirichlet conditions are often encountered in practice, 
e.g. turbulent channel flow with opposite walls held at two fixed temperatures. This special case already presents the essential new 
difficulties and, after analyzing it in detail, we shall briefly comment on modifications required for the general mixed case.

\subsection{Stochastic Representation and Fluctuation-Dissipation Relation}\label{StochRep-FDR-Dbc}

\begin{figure}
\centering    
	\includegraphics[width=.7\linewidth]{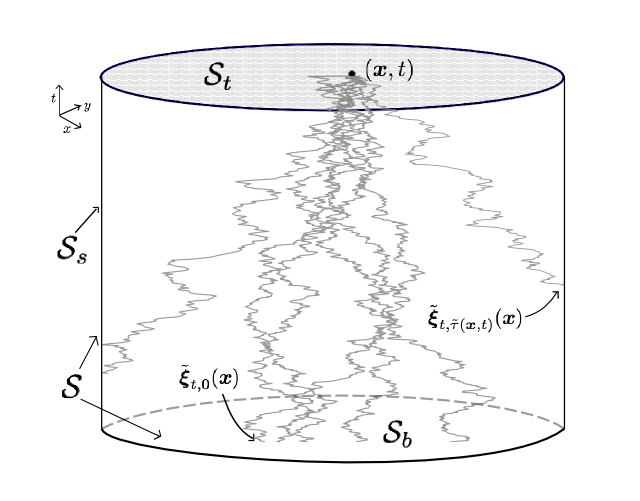}
	\caption{{A schematic of the space-time domain $\cD$ for disk-shaped space domain
	$\Omega\subseteq \mathbb{R}^2$, together with realizations of the process $\tbxi_{t,s}(\bx)$ satisfying 
	Eq. \eqref{stochflowreflected} and stopped at the exit surface $\cS=\cS_b\cup\cS_s$.  Some trajectories 
	reach the base $\cS_b$ at random points $\tbxi_{t,0}(\bx)$ whereas others hit the side-surface $\cS_{s}$ 
	at points $\tbxi_{t,\tilde{\tau}(\bx,t)}(\bx)$ and are stopped there.}}
\label{fig9p5}
\end{figure}

The standard stochastic representation of solutions of the problem (\ref{DBC passive scalar}) involves stochastic particles which are stopped 
at the boundary, contributing a term due to the value maintained at the wall \citep{keanini2007random,soner2007stochastic,oksendal2013stochastic}. 
%A brief review of this theory is in order. 
The stochastic flow may again be defined with reflection at the boundary,  
governed by the equation (\ref{stochflowreflected}) involving the boundary local time. 
The new notion is the {\it boundary (first-)hitting time} or {\it stopping time}, which is defined for $\bx\in\Omega$ by the larger
of $\sup\{s: \tbxi_{t,s}(\bx)\in \partial\Omega\}$, the first time going in reverse to hit the spatial boundary, and the initial time 0, or 
\begin{equation} \tilde{\tau}(\bx,t) = \max\left\{\sup\{s: \tbxi_{t,s}(\bx)\in \partial\Omega\},0\right\}, \end{equation}
%where the backward stochastic flow $\tbxi_{t,s}(\bx)$ is now taken to satisfy 
% \begin{align}\label{stochflowextended}
%\hd \tbxi_{t,s}(\bx)&=\bar{\bu}^\nu( \tbxi_{t,s}(\bx),s)\ \rmd s+\sqrt{2\kappa}\ \hat{\rmd}\bW_s
%\end{align}
%in all of space, with the continuously extended velocity field  
%$$\bar{\bu}^\nu(\bx,t)=\left\{\begin{array}{cl}
%                                      \bu^\nu(\bx,t) & \bx\in\Omega   \cr
%                                      {\bf 0} & \bx\notin \Omega
%                                      \end{array} \right.  . $$
This is the first time (going backward) at which the stochastic Lagrangian particle hits the {space-time \emph{exit surface}} $\cS,$ 
%which is 
the union of the ``side surface'' $\cS_{s}=\partial\Omega\times [0,t)$ and the 
``base'' $\cS_b=\Omega\times \{0\}$ of the right cylindrical domain $\cD=\Omega\times [0,t)$ in 
$(d+1)$-dimensional  space-time, when starting at a point $(\bx,t)$ in the ``top''  $\cS_t=\Omega\times \{t\}.$  {See Figure \ref{fig9p5}.}

To state conveniently the stochastic representation, we define the function which gives the data on $\cS$:
\begin{align}\label{Theta}
\Theta(\bx,s) = \begin{cases} \psi(\bx,s),  & s>0 \ \text{and} \ \bx \in \partial \Omega \\
\theta_0(\bx),  & s=0 \ \text{and} \ \bx \in \Omega   
\end{cases}.
\end{align}
Using the backward It$\bar{{\rm o}}$  formula (\ref{back-Ito-refl}) and integrating from time $t$ down to the hitting
time $\tilde{\tau}(\bx,t)$ gives
\begin{align} \nonumber
\theta(\bx,t) &= \Theta(\tbxi_{t,\tilde{\tau}(\bx,t)}(\bx),\tilde{\tau}(\bx,t))+  \int_{\tilde{\tau}(\bx,t)}^t \rmd t'\ S(\tbxi_{t,t'}(\bx) ,t') \\
& \ \ \ + \sqrt{2\kappa} \int_{\tilde{\tau}(\bx,t)}^t\hd \bW_{t'} \cdot \nabla\theta(\tbxi_{t,t'}(\bx) ,t'), \label{ItoBounded}
\end{align}
where we used the fact that $\tell_{t,s}(\bx)\equiv 0$ for all $\tilde{\tau}(\bx,t)<s<t$  and also used the 
system (\ref{DBC passive scalar}). Taking the expectation over the Brownian motion 
then yields for solutions of the initial-boundary value problem the following stochastic representation: 
\begin{align}\label{DirFK}
\theta(\bx,t) &= \bE\left[ \Theta(\tbxi_{t,\tilde{\tau}(\bx,t)}(\bx),\tilde{\tau}(\bx,t))+  \int_{\tilde{\tau}(\bx,t)}^t \rmd s\ S(\tbxi_{t,s}(\bx) ,s)\right]
\end{align}
See \cite{keanini2007random,soner2007stochastic,oksendal2013stochastic} for more details. 
In particular, the It$\bar{{\rm o}}$  integral in \eqref{ItoBounded} is a (backward) martingale by the optional stopping theorem \citep{oksendal2013stochastic}. 
This formula represents the solution $\theta(\bx,t)$ to the scalar advection-diffusion equation 
as an average over randomly sampled sources and initial-boundary data.  

It is now straightforward to obtain {our first version of a} fluctuation-dissipation relation {by mimicking 
previous arguments}. Applying the It$\bar{{\rm o}}$  isometry valid with the 
random stopping time as the lower range of the integral (e.g. see \cite{oksendal2013stochastic}, Theorem 7.4.1) 
 \begin{align}
\mathbb{E}\left|  \int_{\tilde{\tau}(\bx,t)}^t\hd \bW_{s} \cdot \nabla\theta(\tbxi_{t,s}(\bx) ,s) \right|^2
&=  \bE\left[\int_{\tilde{\tau}(\bx,t)}^t \rmd s\ | \nabla\theta(\tbxi_{t,s}(\bx) ,s) |^2\right]
\end{align}
Thus we obtain from (\ref{ItoBounded}), (\ref{DirFK}) that 
 \begin{align}
 & \var\left[\Theta(\tbxi_{t,
\tilde{\tau}(\bx,t)}(\bx),\tilde{\tau}(\bx,t))+ \int_{\tilde{\tau}(\bx,t)}^t \rmd s\ S(\tbxi_{t,s}(\bx) ,s)  \right]  =
2\kappa\ \bE\left[\int_{\tilde{\tau}(\bx,t)}^t \rmd s\ | \nabla\theta(\tbxi_{t,s}(\bx) ,s) |^2\right]. \label{varbounded-eq}
\end{align}
Finally, averaging over the space domain yields 
\begin{align}\nonumber
 & \frac{1}{2}\left\langle\var\left[\Theta(\tbxi_{t,
\tilde{\tau}(t)},\tilde{\tau}(t))+ \int_{\tilde{\tau}(t)}^t \rmd s\ S(\tbxi_{t,s},s)  \right] \right\rangle_\Omega\\
& \ \ \ \ \ \ \ \ \ \ \ \ \ \ \ \ \ \ \ \ \ \ \ \ \ \ \ \ \ \ =
\kappa\ \left\langle\bE\left[\int_{\tilde{\tau}(t)}^t \rmd s\ | \nabla\theta(\tbxi_{t,s},s) |^2\right]\right\rangle_\Omega. \label{varbounded-ineq}
\end{align}
This exact result is one possible version of a fluctuation-dissipation relation for scalar turbulence with general Dirichlet boundary conditions.

Unlike our previous relations, however,  the righthand side of (\ref{varbounded-ineq}) above is not the total time-integrated scalar 
dissipation over the entire domain of the flow. The relation easily yields a lower bound on the total scalar dissipation
by simply extending the time integration down to $s=0,$ 
\begin{align}
 & \frac{1}{2}\left\langle\var\left[\Theta(\tbxi_{t,
\tilde{\tau}(t)}(\tilde{\tau}(t))+ \int_{\tilde{\tau}(t)}^t \rmd s\ S(\tbxi_{t,s},s)  \right] \right\rangle_\Omega \leq 
\kappa\ \left\langle\bE\left[\int_{0}^t \rmd s\ | \nabla\theta(s) |^2\right]\right\rangle_\Omega. \label{varbounded}
\end{align}
because the $s$-integrand is non-negative and the reflected stochastic flow is volume-preserving{,
so that the Jacobian of the change of variables from $\tbxi_{t,s}(\bx)$ to $\bx$ is unity}. The difference between 
the righthand and lefthand sides is 
\begin{eqnarray}
\Delta^{\nu,\kappa}(t)& \equiv &  
\kappa \int_{0}^t \rmd s\ \frac{1}{|\Omega|}\int_\Omega d^dx\  | \nabla\theta(\bx ,s) |^2- 
\kappa\ \bE\left[\frac{1}{|\Omega|} \int_\Omega d^dx \int_{\tilde{\tau}(\bx,t)}^t \rmd s\  | \nabla\theta(\tbxi_{t,s}(\bx) ,s) |^2\right] \cr 
&=& \kappa \int_{0}^t \rmd s \ \bE\left[ \frac{1}{|\Omega|}\int_{\tilde{\Omega}_{t,s}} d^dx \  | \nabla\theta(\tbxi_{t,s}(\bx) ,s) |^2\right] 
\label{Delta-def} \end{eqnarray}
where 
\begin{equation} \tilde{\Omega}_{t,s} = \{\bx\in\Omega:\ \tilde{\tau}(\bx,t)>s\} \end{equation}
is the set of positions $\bx$ for which the stochastic particle has already hit the boundary by time $s$ (going backward). The inequality 
(\ref{varbounded}) thus fails to be an equality because of the missing contributions to total dissipation at positions of reflected particles. 
We shall discuss further below the sharpness of the inequality (\ref{varbounded}) and whether it should, in a suitable limit, become equality.

\subsection{Numerical Results}\label{DNS-Dbc}

To gain further insight, we present now some numerical results on PDF's of hitting-times for the channel-flow database.  
We considered two release points in the buffer layer at height $\Delta y=10\delta_\tau$ above the bottom wall and at the 
final time $t_f$ of the database. The wall-parallel positions of the release points are those shown in Figure \ref{fig2}, and thus 
one is in a high-speed streak and the other in a low-speed streak. {It turns out that to calculate hitting-time PDF's 
accurately using $N$-sample ensembles of particles is quite difficult, because most particles take a very long time to first
hit the boundary. For our results presented here, } we evolved 
$N=14,336$ samples of stochastic particles solving Eq.~\eqref{stochflowreflected} for three Prandtl numbers $Pr=$0.1, 1, 10
at both release points.  Note that hitting-times $\tau$ satisfying $t_f-\tau\ll \tau_\nu$ and $t_f-\tau\gg \tau_\nu$ are both very rare events 
which could not be observed even with so many samples. We thus consider here the logarithmic variable $\lambda\equiv\ln((t_f-\tau)/ \tau_\nu)$ 
which is appropriate to typical values and calculate PDF's $p(\lambda|\bx,t_f)$. 
The numerical procedures are discussed at length in Appendix \ref{HittingTimeAppendix}. Here we just note that the PDF's
are obtained up to the largest available $\lambda=\ln(t_f/\tau_\nu)\doteq 7.17$ and down to $\lambda=-1,$ but the PDF's at the 
two extremes contain further errors that are not indicated by the error bars (representing both s.e.m. for $N$-samples averages
and variation with kernel bandwidth). The PDF's for $\lambda<1.6$ are shifted to the right by about $5\%$,
because our time-step $\Delta s\doteq 2\times 10^{-3}$ cannot fully resolve the smaller hitting times. For $\lambda>6.5$
the PDF's from kernel-density estimates are too small, because of the endpoint effect due to unavailability of samples for 
$\lambda>7.17.$ Our numerical procedures show the same deficiencies when applied to pure Brownian motion, but 
successfully recover the known analytical results for that case in the range $1.6<\lambda<6.5$.

%NUMERICAL RESULTS HERE (SHOW TAKES LONGER TO HIT AS $\kappa\rightarrow 0,$ WITH $\nu$ FIXED).  

\begin{figure*}
  \begin{subfigure}[b]{1\linewidth}
    \centering
    \includegraphics[width=.45\columnwidth,  height=.45\columnwidth]{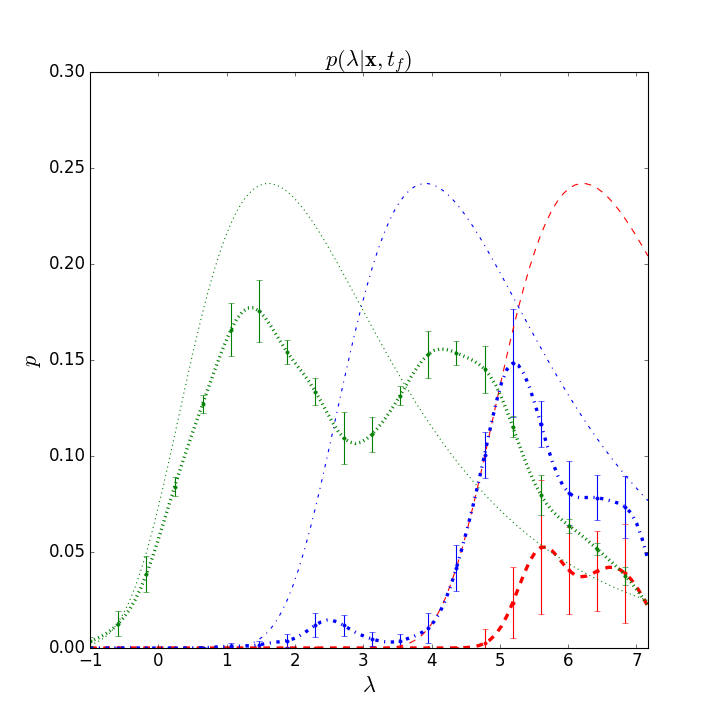} 
    \includegraphics[width=.45\columnwidth,  height=.45\columnwidth]{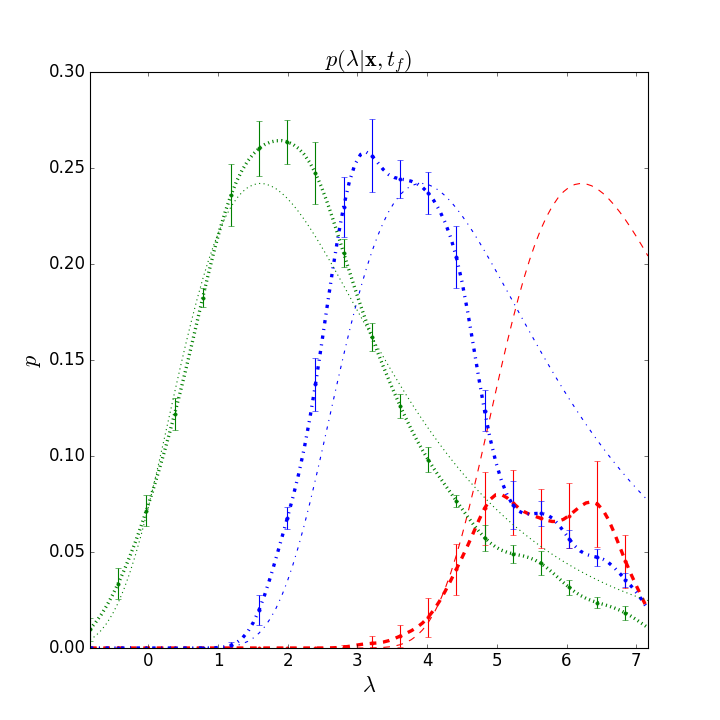} 
  \end{subfigure}  
	\caption{PDF's of the logarithmic hitting-time variable $\lambda=\ln((t_f-\tau)/\tau_\nu).$ 
	\emph{Left panel}: Particles released in a high-speed streak, \emph{Right panel}: particles 
	released in a low-speed streak, for channel flow (heavy) and pure diffusion (light) for Prandtl numbers $Pr = 0.1$ (\textcolor{ForestGreen}{green}, dot, \dottedrule), $1.0$ (\textcolor{blue}{blue}, dash-dot, \dasheddottedrule) and $10$ (\textcolor{red}{red}, dash, \dashedrule).}
\label{fig5}
\end{figure*}

We plot in Fig.~\ref{fig5} the PDF's $p(\lambda|\bx,t_f)$ for the two choices of $\bx,$ the left panel corresponding 
to particles released in a high-speed streak and the right to particles released in a low-speed streak. We also plot for comparison 
the analytical results for hitting-time PDF's with pure diffusion (see eqn. \eqref{hittingtimePDF2} of Appendix \ref{HittingTimeAppendix}).  
One can see in Fig.~\ref{fig5} a very strong effect of the release point on the hitting-time statistics, unlike the situation for the 
boundary local-times accumulated over one flow-through time. At $Pr=0.1$ and $1,$ hitting-times are clearly larger than those of pure 
diffusion for release in a high-speed streak and smaller for release in a low-speed streak. This effect is easy to understand, because  particles 
released in a low-speed streak are advected toward the wall backward in time, and those released in a high-speed streak swept away
from the wall. This effect is not observed for $Pr=10,$ with hitting times for pure diffusion very obviously shorter than those 
for advected particles started in both high-speed and low-speed streaks. This occurs presumably because advection alone 
can never bring a particle to the wall {in a finite time} (because of the vanishing velocity there) and as diffusivity $\kappa$ decreases one 
sees more strongly the effect of fluid advection, which transports the particles away from the wall.    
One can see in general a strong effect of diffusivity on the hitting times, with smaller values of $\kappa$ (or larger $Pr$) leading 
to larger hitting times backward in time. We shall refer to this property in our discussion below of the sharpness of the   
inequality Eq.~\eqref{varbounded}.  However, we first exploit this inequality to prove also for Dirichlet b.c.  that
spontaneous stochasticity is sufficient for anomalous dissipation of passive scalars.

\subsection{Spontaneous Stochasticity Implies Anomalous Dissipation}\label{sec:SS-FDR-Dbc}

The argument is very similar to those given earlier. To present it first in the simplest context, 
we consider a decaying scalar with vanishing source ($S=0$). We first introduce an analogue of the formula 
(\ref{Var2}) for the variance which constitutes the lower bound. There is a natural measure on the {exit surface} $\cS$ ({see Fig. \ref{fig9p5}}) which is 
just the $d$-dimensional Hausdorff 
measure $H^d$ on subsets of $(d+1)$-dimensional spacetime, restricted to $\cS.$ This can be described in 
elementary terms using natural $d$-dimensional coordinates $\bsigma$ on $\cS,$ where on $\cS_b$ we have
$\bsigma=(\by,0)$ for $\by\in \Omega$  and on $\cS_s$ we have $\bsigma=(\bz,\tau)$ for $\bz\in\partial\Omega,$
the presumed smooth bounding wall surface, and $\tau\in [0,t].$ Then for $\bsigma\in \cS$
\begin{equation}  H^d(d\bsigma) =\left\{\begin{array}{ll}
                                        d^dy & \bsigma=(\by,0)\in \cS_b \cr
                                        dH^{d-1}(\bz) d\tau & \bsigma=(\bz,\tau)\in \cS_s 
                                        \end{array} \right. \end{equation}
Here $H^{d-1}$ is the $(d-1)$-dimensional Hausdorff measure on the smooth wall surface $\partial\Omega$ 
(that is, the usual $(d-1)$-dimensional surface area $dS$). We can then also define the delta distribution 
$\delta_\cS(\bsigma',\bsigma)$ on $\cS$ wth respect to measure $H^d,$ so that 
\begin{equation}  \int_\cS H^d(d\bsigma') f(\bsigma')\delta_\cS(\bsigma',\bsigma)=f(\bsigma) \end{equation}
for all smooth functions on $\cS,$ and we can likewise decompose this delta function as 
\begin{equation}  \delta_\cS(\bsigma',\bsigma)=\left\{\begin{array}{ll}
                                        \delta^d(\by'-\by) & \bsigma',\bsigma\in \cS_b \cr
                                        \delta_{\partial\Omega}(\bz',\bz) \delta(\tau'-\tau) & \bsigma',\bsigma \in \cS_s 
                                        \end{array} \right. \end{equation}
Denoting by $\tilde{\bsigma}(\bx,t)=(\tbxi_{t,\tilde{\tau}(\bx,t)}(\bx),\tilde{\tau}(\bx,t))$ the space-time point 
where the trajectory first hits $\cS$ (going backward in time),  we can then introduce 1-time transition probability densities 
for a particle to go backward in time from points 
$(\bx,t)$ on the ``top''  surface $\Omega\times\{t\}$ of the space-time cylinder and to arrive first to $\cS$
at the point $\bsigma\in \cS$:
\begin{eqnarray}
q^{\nu,\kappa}(\bsigma|\bx,t) &=& \bE\left[\delta_\cS\left(\bsigma,\tilde{\bsigma}(\bx,t)\right)\right] \cr 
&=& \left\{\begin{array}{ll}
        \bE\left[\delta^d(\by-\tbxi_{t,0}(\bx))\right] & \bsigma=(\by,0)\in \cS_b \cr
        \bE\left[\delta_{\partial\Omega}(\bz-\tbxi_{t,\tau}(\bx)) \delta(\tau-\tilde{\tau}(\bx,t)) \right] & \bsigma=(\bz,\tau)\in \cS_s 
        \end{array} \right. . \label{transQ}
\end{eqnarray}        
This probability density is normalized with respect to $H^d$ on $\cS$ so that 
\begin{equation}  \int_\cS H^d(d\bsigma) q(\bsigma|\bx,t) =1. \end{equation}
In these terms we obtain a simple formula for the stochastic representation (\ref{DirFK}) in the case of vanishing scalar source:
\begin{equation} \label{hittingTimeRep}
 \theta(\bx,t) = \bE\left[ \Theta(\tilde{\bsigma}(\bx,t))\right]
= \int_\cS H^d(d\bsigma)\  \Theta(\bsigma)\  q(\bsigma|\bx,t). \end{equation}
{We note in passing here that there is another representation for $q(\bsigma|\bx,t)$ in terms of a
process of stochastic Lagrangian particles which is distinct from that considered so far. Rather than letting 
the particles reflect off the wall, one can instead kill the particles when they hit the wall (absorbing boundary conditions). 
If $k(\ba,s|\bx,t)$ for $s<t$ is the transition probability density for this killed process backward in time, then:
\begin{equation}
q(\bsigma|\bx,t) 
=   \left\{\begin{array}{ll}
        k(\by,0|\bx,t) & \bsigma=(\by,0)\in \cS_b \cr
        \bmu(\bz,\tau)\cdot\nabla_\bz k(\bz,\tau|\bx,t) & \bsigma=(\bz,\tau)\in \cS_s 
        \end{array} \right. . 
\label{altDirFK} \end{equation}
Here we make use of the general connection between the joint density function of the hitting time and location and between
the normal derivative of the transition probability for the killed process; see \cite{freidlin1985functional,hsu1986brownian}.
This alternative stochastic interpretation will prove useful for calculations in Appendix \ref{conductionFixedT}.}     

We can obtain a similar formula as (\ref{hittingTimeRep}) for the variance, if we introduce also the corresponding 2-fold transition 
probabilities 
\begin{equation}  q^{\nu,\kappa}_2(\bsigma,\bsigma'|\bx,t) =\bE\left[\delta_\cS\left(\bsigma,\tilde{\bsigma}(\bx,t)\right)
\delta_\cS\left(\bsigma',\tilde{\bsigma}(\bx,t)\right)\right] = \delta_\cS(\bsigma,\bsigma')q(\bsigma|\bx,t), \end{equation}
so that 
\begin{eqnarray}\label{Varbound}
\var\left[\Theta(\tilde{\bsigma}(\bx,t)) \right]  
   &=&   \int_\cS H^d(\rmd\bsigma) \int_\cS H^d(\rmd\bsigma') \ \Theta(\bsigma) \Theta(\bsigma') \cr
&&\hspace{15pt}    \times \Big[q_2(\bsigma,\bsigma'|\bx,t)-q(\bsigma|\bx,t)q(\bsigma'|\bx,t)\Big].
\end{eqnarray}
%where again $q_2(\bsigma,\bsigma'|\bx,t) = \delta_\cS(\bsigma,\bsigma')q(\bsigma|\bx,t)$ for $\bx=\bx'$.
These results are formally identical to those for the source-less $(S=0)$ scalar in domains without 
boundary or with zero-flux Neumann conditions at the wall. Furthermore, $\cS$ is a compact subset of space-time.
Thus, we can exactly mimic our previous arguments with no change. Limits $q^*,q_2^*$ as 
$\nu_j,\kappa_j\rightarrow 0$ always exist along suitable subsequences $\nu_j,\kappa_j$. If the limits were 
deterministic, then one would have
\begin{equation}  q^*_2(\bsigma,\bsigma'|\bx,t) = q^*(\bsigma|\bx,t)q^*(\bsigma'|\bx,t) \label{q2-fac} \end{equation} 
so that $q_2^*$ would factorize in a product of two $q^*$'s. However, if the limits are non-deterministic
so that $q_2^*$ does not factorize for a positive-measure set of $\bx$ at time $t$, then there must be some 
smooth choice of $\Theta=(\theta_0,\psi)$ that makes the non-negative space-averaged variance in (\ref{Varbound}) 
in fact non-zero. For such choice of initial condition $\theta_0$ and boundary conditions $\psi$, the space-average 
and time-integrated scalar dissipation will then be non-zero taking the limit $\nu_j,\kappa_j\rightarrow 0.$ As before, 
this argument works rigorously only for a passive scalar, not an active one. An additional limitation discussed 
further in Appendix \ref{App:A3} is that we cannot use this argument to prove, for a fixed choice of the boundary data $\psi,$ 
that spontaneous stochasticity implies there exists a smooth initial datum $\theta_0$ for which the scalar dissipation rate 
is positive\footnote{
These comments have implications for the problem of Rayleigh-B\'enard convection with fixed temperatures at top/bottom 
plates, which is discussed in {paper III}. There is some evidence for Richardson dispersion of Lagrangian 
particles in numerical simulations of turbulent convection \citep{schumacher2008lagrangian}. However, even if   
there were compelling evidence for spontaneous stochasticity associated to this effect, we could not rigorously 
conclude from our results that there is anomalous thermal dissipation. First, the temperature is an active scalar 
in that case and, second, our present arguments do not work with fixed boundary data.}.  Minor changes to these 
arguments are needed to incorporate a bulk source $S$ {and we leave details also 
to Appendix \ref{App:A3}. }

{The previous arguments do not imply that spontaneous stochasticity is required for anomalous 
scalar dissipation with imposed scalar values at the wall. Even if the factorization (\ref{q2-fac}) holds 
and there is no spontaneous stochasticity, the variance in (\ref{Varbound}) provides only a lower bound 
to the scalar dissipation. Hence, the scalar dissipation rate can tend to a non-zero limit even when the 
variance vanishes! This is not just a failure of the proof, but also an indication that 
other physical mechanisms are available to produce anomalous dissipation with Dirichlet boundary 
conditions. The most obvious alternative mechanism is a narrow scalar boundary layer at the walls.} 
{One can imagine a flow with extremely rapid advective mixing 
in the interior and with opposite scalar values $\pm \psi_0$ imposed at two ends, so that large scalar gradients 
of the order $\nabla\theta\sim O(\psi_0/\kappa)$ are achieved in a boundary layer of thickness $\sim O(\kappa)$
and a near-zero scalar amplitude $\theta\simeq 0$ occurs over the bulk of the domain. Such a flow would 
be the simplest that provides anomalous scalar dissipation from a scalar boundary-layer mechanism.  
For example, one might consider a fluid undergoing solid-body rotation with frequency $\omega$ in the interior 
of the domain and an imposed inhomogeneous temperature distribution on the boundary varying between $-\psi_0$
and $+\psi_0.$ In the rotating frame of reference, this would correspond to a motionless fluid (pure heat conduction)
but with a time-dependent boundary temperature oscillating with frequency $\omega$ between 
the extreme values $\pm \psi_0.$}

{As the simplest example of this type, consider pure diffusion on the interval $[0,H]$ 
with $\theta_0=S\equiv 0$ and boundary temperatures specified as $\psi(0,t)=\psi_0\sin(\omega t)$ and $\psi(H,t)=0.$ 
In this case, the stochastic scalar field is simply
\begin{equation}  
\tth(\bx,t) =   \psi_0\sin(\omega \tilde{\tau}(x,t)).
\end{equation}
The scalar field $\theta(x,t)=\bE[\tth(\bx,t)]$ can be constructed with the knowledge of the hitting-time distribution 
at zero for Brownian motion on the half-line, which is: 
 \begin{align}\label{hittingtimePDF}
p^\kappa(\tau|x,t) = 
\begin{cases} 
 \frac{x}{\sqrt{4\pi\kappa (t- \tau)^3}}\exp\left(-\frac{x^2}{4\kappa (t-\tau)}\right)   &\tau<t\\
0 & \tau> t 
\end{cases}.
\end{align}
See Appendix \ref{conductionFixedT}.  The solution $\theta(x,t)$ of the conduction problem can be obtained by integration over the above distribution, 
and scalar gradients can then be explicitly computed; see \eqref{fixedTDtemp}.  Temperature fluctuations are found to propagate with velocity 
$(\kappa \omega)^{1/2}$ and travel to distances $(\kappa/\omega)^{1/2}$ away from the wall.  In this small boundary layer region, 
the scalar field gradients diverge as $\partial_x \theta \sim \psi_0(\omega/\kappa)^{1/2}$ as $\kappa\rightarrow 0.$ We obtain:
\begin{equation}\label{dimAnalFixedT}
\int_0^L\!\!\!\!  dx \  \kappa |\partial_x\theta(x,t)|^2 \propto (\kappa/\omega)^{1/2} \cdot\kappa\cdot \psi_0^2(\omega/\kappa)\sim \psi_0^2\sqrt{\kappa \omega}. 
\end{equation}
See \eqref{SolAsymGlobalDissFixedT} for a more precise statement.  The scalar field in this simple example exhibits anomalous dissipation 
if $\omega$ is very large for $\kappa\rightarrow 0,$ e.g. with $\omega\sim \kappa^{-\alpha}$ for $\alpha\geq1$.  
In particular, for $\alpha=1$ the limiting scalar dissipation rate is non-zero and finite as $\kappa\rightarrow 0,$ with scalar gradients and 
boundary layer thicknesses as suggested in the previous paragraph.  Thus, just as for flux b.c., thin scalar boundary 
layers can be a source of non-zero (possibly diverging) dissipation as $\nu,\kappa\rightarrow 0$.}

\subsection{Equality Fluctuation-Dissipation Relation}\label{FDR-equal-Dbc}  

We now return to the question whether our lower bound (\ref{varbounded}) on the cumulative (time- and space-integrated)
scalar dissipation may in fact be an equality. This is certainly not true in the $t\rightarrow \infty$ limit with $\kappa,\nu$ fixed. 
We note that 
\begin{equation}  \lim_{t\rightarrow\infty} \tilde{\Omega}_{t,s}=\Omega, \mbox{ any fixed $s>0$} \end{equation}
since the boundary hitting times are finite almost surely for $\kappa,\nu>0$ (see \cite{karlin1981second}, 
Ch.15, section 11) and, thus, the particle released in the far distant future $t$ will certainly hit the space boundary $\partial\Omega$ 
before $s$ (moving backward in time). Thence by dominated convergence
\begin{equation}  \lim_{t\rightarrow \infty}
\bE\left[ \int_{\tbxi_{t,s}\left(\tilde{\Omega}_{t,s}\right)}\!\!\! d^dx\  \kappa\ | \nabla\theta(\bx ,s) |^2\right]
= \kappa\ \int_\Omega d^dx\  | \nabla\theta(\bx ,s) |^2 \end{equation}
and the contribution of reflected particles coincides at long times with the total dissipation.
A strengthened conclusion is that
\begin{equation}  \lim_{t\rightarrow\infty} \tilde{\Omega}_{t,s}=\Omega,  \mbox{ uniformly for all $s$ such that $t-\delta(t)\geq s\geq 0,$} \end{equation}
for some function $\delta(t)\geq 0$ such that 
$ \lim_{t\rightarrow\infty}\delta(t)=\infty, \quad \lim_{t\rightarrow\infty}\frac{\delta(t)}{t}=0. $
This is true because the difference $t-s\geq \delta(t)$ for all such $s.$ In that case, however, we see that
\begin{equation}  \lim_{t\rightarrow\infty} \frac{1}{t}\Delta^{\nu,\kappa}(t)=\lim_{t\rightarrow\infty} \frac{1}{t}\int_0^t ds\ 
 \frac{1}{|\Omega|} \int_\Omega d^dx\  \kappa\ | \nabla\theta(\bx ,s) |^2,\end{equation}
 where $\Delta^{\nu,\kappa}(t)$ is the difference in eq.~(\ref{Delta-def}).  Thus, our lower bound (\ref{varbounded}) on space-time 
average dissipation becomes vacuous as $t\rightarrow\infty$, with the long-time limit of the lower bound simply tending
to zero as reflected particles fill the entire domain.   
 
Now consider the opposite limit with $\nu,\kappa\rightarrow 0$ for fixed $t.$ If we keep $\nu$ fixed so that the velocity 
stays smooth, then no particles will reach the boundary in the finite time interval $[0,t]$ as $\kappa\rightarrow 0$. 
An increase of hitting times with decreasing $\kappa$ for fixed $\nu$ is clearly observed in Fig.~\ref{fig5}. In this 
limit, the inequality in our lower bound becomes equality. However, as long as $\bu$ is smooth, then there will be no dissipative anomaly and the scalar 
dissipation will vanish! Thus, this limit is of little interest. On the other hand, if we take $\nu,\kappa\rightarrow 0$ together, then it is 
possible that the particles will hit the boundary and reflect. In fact, this must be the case if there is a scalar dissipative anomaly 
at finite times $t$ for flux b.c. and vanishing initial data of the scalar and vanishing scalar sources, since then the FDR (\ref{FDR-Nbc}) 
is solely due to the contribution of the boundary local times in (\ref{tth-Nbc}). If particles continue to hit the boundary as  
$\nu,\kappa\rightarrow 0$ for fixed time $t,$ then our lower bound must be a strict inequality (but possibly non-vanishing
for finite $t$). 

It is straightforward, however, to obtain an equality relation for the scalar dissipation by using the same derivation 
as for flux-b.c, namely, by using the backwards It$\bar{{\rm o}}$  formula (\ref{back-Ito-refl}) and integrating from time 
$t$ down to $0,$ rather than stopping when hitting the boundary. This argument yields for the scalar itself
\begin{equation}
\theta(\bx,t)=\bE\left[  \theta_0(\tbxi_{t,0}(\bx))
+ \int_{0}^t \rmd s \ S{(\tbxi_{t,s}(\bx),s) }  - \kappa\int_{0}^t {{\boldsymbol{\mu}}\cdot \nabla\theta(\tbxi_{t,s}(\bx),s)   \ \hd \tell_{t,s}}(\bx)\right] 
\label{theta-Dbc}
\end{equation}
and for the variance
\begin{align}\nonumber
\frac{1}{2} \left\langle\var\left[  \theta_0(\tbxi_{t,0})
+ \int_{0}^t \rmd s \ S{(\tbxi_{t,s},s) }  - \kappa\int_{0}^t {{\boldsymbol{\mu}}\cdot \nabla\theta(\tbxi_{t,s}(\bx),s)   
\ \hd \tell_{t,s}}\right] \right\rangle_\Omega&\\
\  \  \  \ \ \ \ \ = \kappa \int_{0}^t \rmd s \left\langle | \nabla\theta(s) |^2\right\rangle_\Omega, & \label{NeumannVar}
\end{align}
which is {our second} version of a fluctuation-dissipation relation for Dirichlet b.c. of the scalar. The total scalar dissipation is 
now obtained, with the contribution of the reflected particles represented by the boundary local time density term on the left. 
Unfortunately,  this new formula is not very convenient for mathematical analysis, because 
the scalar boundary flux $g(\bx,t)=-{{\boldsymbol{\mu}}}\cdot\nabla\theta(\bx,t)$ is no longer a known input, but is 
instead a space-time fluctuating quantity which must be determined from the solution of the scalar advection-diffusion equation.   
{This relation thus has a mixed Euler-Lagrangian character, because it involves both the Eulerian scalar field
$\theta(\bx,t)$ and the stochastic Lagrangian flow $\tbxi_{t,s}.$ Despite being clumsy for mathematical use, 
this relation} is the physically most natural version of the FDR for  scalar Dirichlet b.c., and provides insight into 
the Lagrangian origin of scalar dissipation. {In Appendix \ref{altFeynKacAppend}, we explicitly demonstrate the 
equivalence of the formula \eqref{theta-Dbc} to the more standard representation \eqref{hittingTimeRep}
for the simple example of heat conduction with oscillating wall temperature.}

\subsection{Mixed Boundary Conditions}\label{mixed}

The extension of the previous results to scalars with general mixed Dirichlet-Neumann conditions (\ref{mixed-bc-D}),(\ref{mixed-bc-N})
is straightforward. Again, one considers the backward stochastic flow with reflection at the boundary. To derive an FDR 
in the form of an inequality like (\ref{varbounded-ineq}), one must stop those particles which hit $\partial\Omega_D$ before time $0$ 
(going backward). The particles which hit $\partial\Omega_N$ are simply reflected. Results like (\ref{varbounded-eq}) and (\ref{varbounded-ineq})
are obtained, except that now the stopping time $\tilde{\tau}(\bx,t)$ is the first time to hit $\partial\Omega_D$ (or 0, whichever is larger)
and the variance on the left includes contributions from the local time density at the piece $\partial\Omega_N$ of the boundary. 
This FDR inequality omits the contribution to scalar dissipation at locations of particles reflected from $\partial\Omega_D$. It is also easy to 
obtain an FDR equality like (\ref{NeumannVar}) by stopping no particles.  It has an identical form to (\ref{NeumannVar}) 
and the sole difference is that $g(\bx,t)=-{{\boldsymbol{\mu}}}\cdot\nabla\theta(\bx,t)$ is a specified function for points $\bx\in \partial\Omega_N$
but must be obtained for $\bx\in \partial\Omega_D$ by solving the advection-diffusion equation. {We shall make use of this 
version of the FDR in our discussion of turbulent Rayleigh-B\'enard convection in the following paper III}.

\section{Summary and Discussion}\label{sec:summary}
 
 {This paper has extended to flows in wall-bounded domains the Lagrangian fluctuation-dissipation relation 
 introduced in Paper I for scalars advected by an incompressible fluid. 
 This relation expresses an exact balance between  molecular dissipation of the scalar and input of scalar variance 
 from the initial values, boundary fluxes, and internal sources as these are sampled by stochastic Lagrangian trajectories 
 backward in time.  We have exploited this relation to prove, for domains with no scalar flux through the wall, that spontaneous stochasticity 
 of Lagrangian trajectories is necessary and sufficient for  anomalous dissipation of passive scalars, and necessary 
 (but possibly not sufficient) for anomalous dissipation of active scalars.  Cream stirred into coffee is an everyday example 
 of this type, as is any scalar advected by a fluid in a container with impermeable walls. 
 For more general mixed boundary conditions on the scalar, with imposed values at the wall or imposed non-zero fluxes, 
 simple  examples show that thin scalar boundary layers provide a distinct mechanism for non-vanishing scalar dissipation. 
 Nevertheless, we can still show rigorously for general scalar boundary conditions that spontaneous stochasticity is sufficient 
 for anomalous dissipation of passive scalars, and this result plausibly extends to active scalars as well. Thus, in addition 
 to scalar boundary layers, Lagrangian spontaneous stochasticity is shown here to be another possible source of 
 anomalous scalar dissipation in wall-bounded flows.}  

{An interesting issue for further study is whether Lagrangian spontaneous stochasticity plays any role
in anomalous dissipation of kinetic energy in wall-bounded flows. There is some evidence from 
experimental measurements of turbulence in closed containers that, while the kinetic energy dissipation
due to viscous boundary layers decreases very slowly with increasing Reynolds number, the energy 
dissipation in the bulk of the turbulent flow is very nearly Reynolds-number independent \citep{cadot1997energy}.
It appears possible that such ``anomalous dissipation'' in the bulk is produced by very similar mechanisms 
as energy dissipation in homogeneous, isotropic turbulence and thus may be related to spontaneous 
stochasticity in the bulk, as discussed in paper I. Note that the stochastic Lagrangian representation 
of incompressible Navier-Stokes solutions discussed in paper I has been successfully extended to wall-bounded 
flows with stick boundary conditions for the velocity at the wall \citep{ConstantinIyer11}. The Lagrangian
dynamics of vorticity in this stochastic formulation generalizes the classical Helmholtz theorem for ideal 
Euler solutions. Indeed, similar to the case of scalar fields with Dirichlet conditions discussed in section \ref{Dirichlet},
the vorticity field at a point is an average of ``frozen-in'' vorticity vectors that are transported along an ensemble 
of stochastic Lagrangian trajectories backward in time. Those trajectories that hit the wall transport the vorticity 
that they encounter there, while those that never hit the wall before time 0 transport the initial vorticity. 
See \cite{ConstantinIyer11} for detailed discussion and proofs. These results make it possible to study in detail 
the contribution of Lagrangian vorticity dynamics, e.g. stretching of vortex filaments, to turbulent energy
dissipation in wall-bounded flows. A crucial difference from homogeneous, isotropic turbulence, however, 
is that random stretching dynamics is insufficient to explain energy dissipation in the presence of walls,
but instead vortex lines generated at the wall must undergo organized motion away from the wall 
\citep{taylor1932transport,huggins1970energy,eyink2008turbulent}. We therefore expect that the 
effects of walls on Lagrangian mechanisms of turbulent energy dissipation are quite profound.} 

The Lagrangian fluctuation-dissipation relation proved in the present paper is valid in wall-bounded 
flows for all scalars, passive or active, whether either Lagrangian spontaneous stochasticity
or scalar anomalous dissipation occur or not in a particular flow. In general, the FDR provides a novel Lagrangian 
view of turbulent scalar dissipation. {In the following paper III we apply the FDR derived here to turbulent  
Rayleigh-B\'enard convection, where the mean thermal dissipation rate
and the mean kinetic energy dissipation rate are closely related to each other and also to the vertical 
heat transport through the convection cell. In Rayleigh-B\'enard turbulence it is well-known that 
dissipative anomalies for kinetic energy and thermal fluctuations, if they exist, lead to an ``ultimate regime'' 
of convection at high Rayleigh numbers, with scaling of the heat flux as predicted by 
\cite{spiegel1971convection} or, with a logarithmic correction, by \cite{kraichnan1962turbulent}. 
The absence of dissipative anomalies instead leads 
to a weaker dependence of heat flux on Rayleigh number than predicted by the Kraichnan-Spiegel theories.
Exploiting our Lagrangian FDR's \eqref{FDR-Nbc} and \eqref{NeumannVar} we are able to relate the thermal
dissipation rate in Rayleigh-B\'enard turbulence to the integral mixing time required for passive tracers released 
at the top or bottom wall to attain to their final uniform value near those walls. We conclude from our analysis that 
dissipative anomalies and Kraichnan-Spiegel scaling will hold, unless this near-wall mixing time is asymptotically 
much longer than the gravitational free-fall time or, nearly, the large-scale circulation time.}  This concrete application 
thus provides a wider window into the Lagrangian mechanisms of turbulent dissipation.

% \newpage 
 
 \section*{Acknowledgements}

We would like to thank Navid Constantinou, Cristian C. Lalescu and Perry Johnson for useful 
discussions.  We would like to thank the Institute for Pure and Applied Mathematics (IPAM) at UCLA, 
where this paper was partially completed during the fall 2014 long program on ``Mathematics 
of Turbulence''. We also acknowledge the Johns Hopkins Turbulence Database for the numerical 
turbulence data employed in this work. G.E. is partially supported by a grant from
NSF CBET-1507469 and T.D. was partially supported by the Duncan Fund and a Fink Award 
from the Department of Applied Mathematics \& Statistics at the Johns Hopkins University.

 \appendix

%\section{Prior Work in the Kraichnan Model (and Beyond)}\label{BGK}
 
\section{{Heat Conduction with Imposed Fluxes}}\label{conduction}

{As an illustration of the general formalism and also as basis of comparison for advected
scalars, we consider here the Neumann problem for the diffusion equation on the interval $[0,H],$  
with imposed fluxes at the endpoints. As in the main text, we adopt the terminology of 
heat conduction and thus consider the scalar to be temperature.} 
We work out explicit analytical results for the stochastic representation 
and fluctuation-dissipation  relation in the limit $\kappa\to 0$ at any fixed time $t.$ 
%In Appendix B of paper III we shall consider the limit $t\to \infty$ at fixed $\kappa$. 

%The problem then becomes: 
{The problem considered is stated precisely as:}
\begin{align}\nonumber
\partial_t T &= \kappa \partial_x^2 T\ \ \  {\rm for} \ \ \  x\in[0,H]\\ \label{pureCond}
\kappa\partial_x T &= - J \ \ \ \ \ \  {\rm at} \ \ \ x=0,H\\ 
 T& = 0 \ \ \ \ \ \  \ \ \   {\rm at} \ \ \ t=0.\nonumber
\end{align}
The scalar flux $J$ at the boundaries is a space-time constant and also independent of $\kappa$.  
The Green's function for the problem \eqref{pureCond} or, equivalently, the transition probability density 
of the reflected Brownian motion, satisfies, for all $t>0$ and $x\in [0,H]$, the backward evolution equation:
\begin{align}\nonumber
\partial_s p(a,s|x,t) &=- \kappa \partial_a^2 p(a,s|x,t)\ \ \  {\rm for} \ \ \  (a,s)\in[0,H]\times [0,t]\\
\partial_a p(a,s|x,t) &= 0 \ \ \  {\rm at} \ \ \ a=0,H \label{pureCondTrans}\\
p(a,t|x,t) &=  \delta(a-x).\nonumber
\end{align}
The solution to \eqref{pureCondTrans} can be represented using Fourier cosine series:
\begin{align}\label{transRep}
p(a,s|x,t) &=  \frac{1}{H} + \frac{2}{H} \sum_{n=1}^\infty \cos\left(\frac{n\pi}{H} x\right) \cos\left(\frac{n\pi}{H} a\right) e^{-\kappa\left(\frac{n\pi}{H}\right)^2 (t-s)}.
\end{align}
%We will investigate ``dissipative anomalies" for the solution of \eqref{pureCond} as $\kappa\to 0$ through our fluctuation-dissipation relations  
%at both finite and infinite time. 

\subsection{Temperature Profile and Boundary Layer}\label{limtemp:finitetime}

The temperature field can be represented via mean boundary local times at $x=0$ and $x=H$ 
using the general stochastic representation formula \eqref{NeumannStochRep}: 
\begin{equation}\label{tempsol}
 T(x,t) = -J \Big( \mathbb{E}\tilde{\ell}_{t,0}^0(x) -\mathbb{E}\tilde{\ell}_{t,0}^H(x) \Big) =J\int_0^tds\ \Big(  p(0,s|x,t) -  p(H,s|x,t) \Big).
\end{equation}
We first study the limiting profile $T(x,t)$ as $\kappa\rightarrow 0$ via the stochastic representation (\ref{tempsol}). 
In the coarsest sense, this limit can be obtained from the convergence of the mean local times to delta-distributions at the boundary.  
Indeed, integrating against an arbitrary smooth test function $\varphi\in C^\infty[0,H]$ satisfying $\varphi'(0)=\varphi'(H)=0,$ we have, 
using the representation \eqref{transRep}:
\begin{align}
-\int_0^Hdx \ \varphi(x) \ \mathbb{E} \tilde{\ell}_{t,0}^{\sigma H} (x)&=  a_0 t + \frac{H^2}{\kappa \pi^2} \sum_{n=1}^\infty  \frac{(-1)^{n\sigma}}{n^2} \left(1-  e^{-\kappa\left(\frac{n\pi}{H}\right)^2 t}\right)a_n, 
\end{align}
where $\sigma=0,1$ and 
\begin{align}
a_0 = \frac{1}{H} \int_0^H \varphi(x)dx,  \ \ \ a_n = \frac{2}{H} \int_0^H \varphi(x) \cos\left(\frac{n\pi}{H} x\right) dx.
\end{align}
Because the Fourier coefficients $a_n$ are rapidly decaying in $n$ for smooth $\varphi,$ the series is absolutely convergent and the $\kappa\to 0$ limit yields:
\begin{align}
\lim_{\kappa\to 0} \int_0^Hdx \ \varphi(x)\ \left(-\mathbb{E} \tilde{\ell}_{t,0}^{\sigma H} (x)\right)&=  t\left[a_0  +  \sum_{n=1}^\infty  (-1)^{n\sigma}a_n\right]= t\varphi(\sigma H)
\end{align}
where we have used $\varphi(x) = a_0 + \sum_{n=1}^\infty a_n\cos\left(\frac{n\pi}{H} x\right)$.  Therefore we find:
\begin{align}\label{convtoDeltas}
-\lim_{\kappa\to 0}\mathbb{E} \left[\tilde{\ell}_{t,0}^0 (x)\right] = t \delta(x),  \ \ \ \ -\lim_{\kappa\to 0}\mathbb{E} \left[\tilde{\ell}_{t,0}^H (x)\right] = t \delta(x-H),  
\end{align}
in the sense of distributions and thus
\begin{align}
 \lim_{\kappa\rightarrow 0}T(x,t) &=J t\Big(\delta(x)- \delta(x-H)\Big) \label{zeroKappaheat}
 \end{align}
in the sense of distributions. 

These delta-functions represent thin thermal boundary layers for small values of $\kappa>0.$ The small-$\kappa$
asymptotics can be inferred by rewriting the transition density \eqref{transRep} as:
\begin{align}\nonumber
p(a,s|x,t) &= \frac{1}{H} \sum_{n=-\infty}^\infty \cos\left(\frac{n\pi}{H} x\right) \cos\left(\frac{n\pi}{H} a\right) e^{-\kappa\left(\frac{n\pi}{H}\right)^2 (t-s)}\\\label{transRep2}
 &=  \frac{1}{\sqrt{4\kappa \pi (t-s)}} \sum_{n=-\infty}^\infty  \left(e^{-\frac{ (x-a+2nH)^2}{4\kappa (t-s)}} + e^{{-\frac{ (x+a+2nH)^2}{4\kappa (t-s)}}}\right).
\end{align}
where we have employed the Poisson summation formula (e.g. see \cite{katznelson2004introduction})
in passing to the second equality.  This is a useful reformulation since series \eqref{transRep2} is rapidly 
convergent for $\kappa (t-s)/H^2\ll 1$ whereas \eqref{transRep} converges rapidly for $\kappa (t-s)/H^2\gg1$.  
Thus, at fixed $t$ and $H$ we can study the zero-diffusion limit $\kappa\to 0$ directly from this formula.  
In particular, since $x,a\in[0,H]$, the asymptotic transition probability is: 
\begin{align}\label{zeroDiffasymTrans}
p(a,s|x,t) \sim  \frac{1}{\sqrt{4\kappa \pi (t-s)}} \left(e^{-\frac{ (x-a)^2}{4\kappa (t-s)}} + e^{-\frac{ (x+a)^2}{4\kappa (t-s)}} 
+ e^{-\frac{ (x+a-2H)^2}{4\kappa (t-s)}} \right)  \  \text{ as }  \ \   \kappa \to 0. 
\end{align}
All other terms in the sum \eqref{transRep2} vanish transcendentally.  In fact, for $0\leq a,x\ll H$ only two terms remain, 
\begin{align}\label{zeroDiffasymTrans2}
p(a,s|x,t) \sim  \frac{1}{\sqrt{4\kappa \pi (t-s)}} \left(e^{-\frac{ (x-a)^2}{4\kappa (t-s)}} + e^{-\frac{ (x+a)^2}{4\kappa (t-s)}} 
 \right)  \  \text{ as }  \ \   \kappa \to 0,
\end{align}
which may be recognized as the transition probability for the diffusion process $\tilde{X}(s)=|x+\sqrt{2\kappa}\tilde{W}(t-s)|$
on the half-line $[0,\infty),$ which is a (scaled) Brownian process starting at $x$ at time $t$ moving backward in time $s$ 
and reflected at the origin.  This result is intuitively clear, since the right boundary of the interval at $H$ should play no role 
for $a,x\ll H$ in the limit $\kappa\rightarrow 0$ at fixed times. By the exact symmetry $p(H-a,s|H-x,t)=p(a,s|x,t),$ there is 
a similar asymptotics with only two terms surviving in (\ref{zeroDiffasymTrans}) for $a,x$ near $H$.  
We shall thus restrict our asymptotic analysis of the thermal boundary-layer to that near $x=0,$ and infer the result near $x=H$
by symmetry. 

We introduce a scaled coordinate $\xi=x/\sqrt{\kappa t}$ which is held fixed as $\kappa\rightarrow 0.$ 
From the mean value $-\mathbb{E}\tilde{\ell}_{t,0}^0(x) =\int_0^tds\ p(0,s|x,t)$ and (\ref{zeroDiffasymTrans2}) one easily obtains that 
$$ -\mathbb{E}\tilde{\ell}_{t,0}^0(\sqrt{\kappa t}\xi)\sim \left(\frac{t}{\kappa}\right)^{1/2} f(\xi), \quad \mbox{ as }
\kappa\rightarrow 0, 
\quad  \mbox { with } f(\xi) = \frac{1}{\sqrt{\pi}} \int_0^1 \frac{du}{\sqrt{u}} \exp\left(-\frac{\xi^2}{4u}\right). $$ 
The scaling function $f(\xi)$ satisfies $f(0)=2/\sqrt{\pi},$ $\int_0^\infty d\xi\ f(\xi)=1,$ and vanishes rapidly for $\xi\gg 1.$ 
%As expected \eqref{zeroDiffasymTrans} reduces to the the transition probability of Brownian motion on the half line $[0,\infty)$ when $x,a$ are distances $\mathcal{O}(\sqrt{\kappa t})$ from the boundary $x=0$:  
%\begin{align}\label{zeroDiffasymTrans2}
% \sqrt{\kappa t} \ p(\sqrt{\kappa t} \ \alpha,s|\sqrt{\kappa t} \ \xi,t) \sim  \frac{1}{\sqrt{4 \pi (1-s/t)}} \left(e^{-\frac{ (\xi-\alpha)^2}{4 (1-s/t)}} + e^{-\frac{ (\xi+\alpha)^2}{4 (1-s/t)}}  \right)  \  \text{ as }  \ \   \kappa \to 0. 
%\end{align}
%for fixed $\xi=x/\sqrt{\kappa t}$ and $\alpha =a/\sqrt{\kappa t}$.  We now analyze the behavior of the mean local times in this ``boundary layer" of width $\mathcal{O}(\sqrt{\kappa t})$ near $x=0$.  Setting $a=\alpha = 0$:
%\begin{align*}
%\sqrt{\frac{\kappa}{t}}\  \mathbb{E} \tilde{\ell}_{t,0}^0 (\sqrt{\kappa t} \ \xi )
%&=   \frac{1}{t} \int_0^t ds  \frac{1}{\sqrt{\pi   (1-s/t)} }\exp\left({-\frac{ \xi^2}{4 (1-s/t)}}  \right)=\mathcal{O}(1).
%\end{align*}
The mean of the boundary local time $\tell_{t,0}^0(x)$ vanishes transcendentally for $x$ in the interior $(0,H)$ of the domain, 
but the convergence is highly non-uniform. In particular, at points within a distance $\sqrt{\kappa t}$ of the walls, 
the mean local times are of order $\sqrt{t/\kappa}$.  We immediately conclude for the temperature field that 
\begin{align}
 T(x,t) &\sim  J  \left(\frac{t}{\kappa}\right)^{1/2} \left[f\left(\frac{x}{\sqrt{\kappa t}}\right) - f\left(\frac{H-x}{\sqrt{\kappa t}}\right)  \right]
 \label{ScalingzeroKappaheat}
 \end{align}
%Using the representation \eqref{tempsol} and the asymptotics \eqref{zeroDiffasymTrans}, we obtain the solution of the \eqref{pureCond} as $\kappa \to 0$:
%\begin{align}
% T(x,t) &\sim  J  \int_0^t ds \ \frac{1}{\sqrt{\kappa \pi (t-s)}} \left[\exp\left({-\frac{ |x|^2}{4\kappa (t-s)}}\right) - \exp\left({-\frac{ |x-H|^2}{4\kappa (t-s)}}\right)\right]\label{ScalingzeroKappaheat}
 %\end{align}
 when $\kappa\ll H^2/t$.  This asymptotics refines the earlier result \eqref{zeroKappaheat}. 
 Without the aid of the diffusivity to spread it, the temperature rises unboundedly at the walls as $\kappa\rightarrow 0$.  
 
 We can use the above asymptotics to study also the temperature gradient and the scalar dissipation. 
 Indeed, by differentiating the formula \eqref{ScalingzeroKappaheat} and using $f'(\xi)=
 -{\rm erfc}(\xi/2)$ in terms of the complementary error function, we obtain 
\begin{align}\label{limTempGrad}
\partial_x T(x,t) \sim -\frac{ J}{\kappa}  \left[{\rm erfc}\left(\frac{x}{\sqrt{4\kappa t}}\right)+ {\rm erfc }\left(\frac{H-x}{\sqrt{4\kappa t}}\right)\right]
 \end{align}
for $\kappa\ll H^2/t$. This clearly satisfies the flux b.c. in (\ref{pureCond}). We can exploit this formula to 
evaluate the space-time average of the thermal dissipation field $\varepsilon_T(x,t)=\kappa|\partial_x T(x,t)|^2$
in the limit $\kappa\rightarrow 0.$ Using the improper integral $\int_0^\infty {\rm erfc}^2(z) \ dz = \frac{1}{\sqrt{\pi}}
(2-\sqrt{2})$ \citep{ng1969table}, it is straightforward to show that 
\begin{align}\label{SolAsymGlobalDiss}
 \langle \varepsilon_T \rangle_{V,t}\equiv \frac{1}{t} \int_0^tds \frac{1}{H}\int_0^Hdx\ \kappa|\partial_x T(x,s)|^2 \sim \frac{8}{3\sqrt{\pi}}\left(2 -\sqrt{2}\right)\frac{J^2}{ H}\sqrt{\frac{t}{\kappa}}. 
 \end{align}
 Apart from the precise numerical prefactor, this is the scaling that one would expect for temperature gradients of 
 magnitude $\partial_x T \sim {J}/{\kappa}$ in a boundary layer of thickness $\sim \sqrt{\kappa t}.$
 %, leading to a ``dissipative anomaly" at finite times of the form:
%\begin{align}\label{finitetimeGlobalDiss}
% \langle \varepsilon_T \rangle_{V,t}  \sim\kappa \left(\frac{J}{\kappa}\right)^2  \frac{\sqrt{\kappa t}}{H} = \frac{J^2}{H}\sqrt{\frac{{t}}{\kappa }}.
%\end{align}

\subsection{Fluctuation-Dissipation Relation} \label{checkNeumannFDR}

{We calculate next the variance in our local FDR (\ref{loc-FDR-Nbc}) for the problem (\ref{pureCond}): 
\begin{equation}
 \langle\varepsilon_T^{fluc}(x)\rangle_{t}=
\frac{J^2}{2t}{\rm Var}\left[ \left(\tell_{t,0}^{L}(x)-\tell_{t,0}^{R}(x)\right) \right] \label{FDR-pureCond} \end{equation}
This ``fluctuational dissipation'' has the same average over space 
as the time-averaged molecular dissipation $\frac{1}{t}\int_0^t ds\ \varepsilon_T(x,s)$ and the two quantities 
are presumably spatially well-correlated at short times. 
To evaluate the space-average of (\ref{FDR-pureCond}), we employ the general formula \eqref{Var-g} to write}
\begin{eqnarray}
 && \langle\varepsilon_T^{fluc}\rangle_{V,t}=  \frac{2J^2}{H}
\int_0^{H} dx \ \frac{1}{t}\int_0^tds\int_s^tds'\ \cr
&& \times \left(\Big[ p(0,s|0,s')- p(0,s|x,t)\Big]  - \Big[  p(H,s|0,s')-  p(H,s|x,t)\Big]  \right)p(0,s'|x,t).
\label{FDR-fint-AppB} \end{eqnarray}
Here we have used the symmetry of the integral in $s,s'$ and the backward Markov property of the transition probabilities.  
%(e.g. see \eqref{VarJ-T2} in the following appendix). 
We have also made use of the $0\leftrightarrow H$ symmetry after integrating over $[0,H]$.  
Using the asymptotics \eqref{zeroDiffasymTrans}, we have for any $x\in [0,H]$, keeping only the leading terms:
\begin{eqnarray}
\left[p(H,s|0,s') - p(H,s|x,s')\right]p(0,s'|x,t) &\sim &\Bigg[\frac{e^{-\frac{ H^2}{4\kappa (t-s)}}}{\sqrt{4\kappa \pi (s'-s)}} - \frac{e^{-\frac{ (x-H)^2}{4\kappa (t-s)}}}{\sqrt{4\kappa \pi (t-s)}}  \Bigg]\Bigg[ \frac{e^{-\frac{ x^2}{4\kappa (t-s')}}}{\sqrt{4\kappa \pi (t-s')}} \Bigg]\cr
&& \longrightarrow 0  \ \  \text{transcendentally,} \ \ \text{ as } \ \ \kappa \to 0 .  
\end{eqnarray} 
Thus, all heterohedral terms vanish as $\kappa\to 0$ and only homohedral terms contribute to the scalar variance. 
Because of our use of the $0\leftrightarrow H$ symmetry, only the contributions from $\langle\varepsilon_T^{fluc}(x)\rangle_t$
for $0\leq x\ll H$ survive (multiplied by 2), where as $\kappa\rightarrow 0$
\begin{eqnarray} 
 \langle\varepsilon_T^{fluc}((\kappa t)^{1/2}\xi)\rangle_t &\sim &\frac{J^2}{t}\nonumber
  \int_0^tds\int_s^tds'\ \Big[ p(0,s|0,s')- p(0,s|\sqrt{\kappa t}\ \xi,t)\Big] p(0,s'|\sqrt{\kappa t} \ \xi,t) \cr
&\sim& \frac{J^2}{\kappa\pi t}  
       \int_0^tds\int_s^tds'\ \left[ \frac{\exp\left(-\frac{\xi^2 t}{4(t-s')}\right)}{\sqrt{(s'-s)(t-s')}}
       -\frac{\exp\left(-\frac{\xi^2 t(2t-s-s')}{4(t-s)(t-s')}\right)}{\sqrt{(t-s)(t-s')}}
       \right]
       \end{eqnarray}
Integrating over $\xi$ from 0 to $\infty$ and multiplying by $2/H$ gives, after a change of variables to $u=s/t,$ $u'=s'/t,$
 \begin{equation}  \langle\varepsilon_T^{fluc}\rangle_{V,t}\sim \frac{2}{\sqrt{\pi}} \frac{J^2}{H} \left(\frac{t}{\kappa}\right)^{1/2}
     \int_0^1du\int_u^1 du' \left[\frac{1}{\sqrt{u'-u}}-\frac{1}{\sqrt{2-(u+u')}}\right]. \end{equation}      
The double integral is easily reduced to elementary integrals, with the value $\frac{4}{3}(2-\sqrt{2}).$ Thus,  we obtain finally      
 \begin{equation}
\langle\varepsilon_T^{fluc}\rangle_{V,t}\sim \frac{8}{3\sqrt{\pi}}\left(2 -\sqrt{2}\right)\frac{J^2}{ H}\sqrt{\frac{t}{\kappa}},
\qquad\kappa\rightarrow 0, 
 \end{equation}    
in exact agreement with \eqref{SolAsymGlobalDiss}.  

\begin{figure}
\centering
	\includegraphics[width=0.7\linewidth,height=0.5\linewidth]{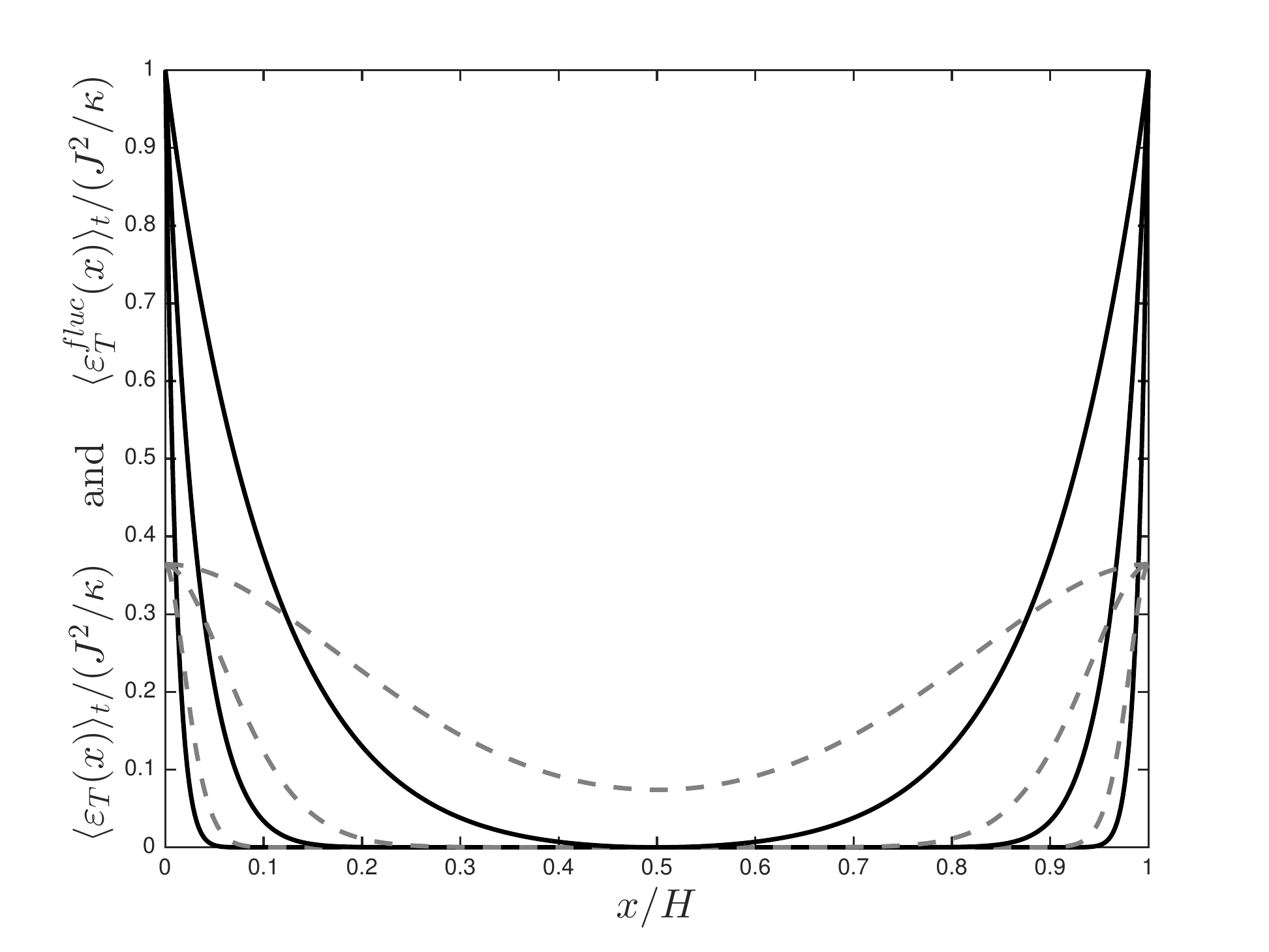}
	\caption{Local thermal dissipation (black, solid) and scalar variance (grey, dashed) for pure heat conduction
	on a 1D interval, non-dimensionalized by $J^2/\kappa$, plotted for three dimensionless times $\tau\equiv \kappa t/H^2=5\times 10^{-2}, \ 5\times 10^{-3}, \ 5\times 10^{-4}.$ The dissipation and variance both localize near the walls as $\tau$ decreases.}
\label{fig:heatDiss}
\end{figure}

The local dissipation measures $\langle\varepsilon_T^{fluc}(x)\rangle_t$ and $\langle \varepsilon_T(x)\rangle_{t}$ 
need not be the same pointwise, but
%, as argued in Section \ref{FDR-RBFB}, 
they should be closely correlated.
To test this idea for pure heat conduction, we have numerically evaluated {the $x$-integrand in (\ref{FDR-fint-AppB})
after re-symmetrization under the reflection $x\leftrightarrow H-x$, by}
using the analytical representation \eqref{transRep2} of the transition probabilities
truncated to a finite number of terms. The double time-integral in \eqref{FDR-fint-AppB} was evaluated using the Matlab
function {\tt integral2}, setting absolute tolerance {\tt abstol=1e-10} and relative tolerance {\tt reltol=1e-8.} In Fig.~\ref{fig:heatDiss}
we plot the resulting local scalar variance $\langle\varepsilon_T^{fluc}(x)\rangle_t$  using the three-term expansion \eqref{zeroDiffasymTrans}, 
but we have verified that adding further terms from the full expansion \eqref{transRep2} made no observable change. 
We also plot for comparison the local dissipation $\langle \varepsilon_T(x)\rangle_{t}$ 
obtained from the asymptotic temperature gradient \eqref{limTempGrad}.  
Though not identical, we see a close qualitative correspondence of the spatial behavior of these two dissipation measures, 
particularly for small dimensionless time $\tau\equiv \kappa t/H^2$ when the dissipation accumulates at the walls.

\subsection{Boundary Local Time Density of Reflected Brownian on the Half-Line}\label{reflBr} 

The formula \eqref{loctim-pdf} invoked in section \ref{sec:SS-AnomDiss-Nbc} is a standard 
result for a Brownian motion on the half-line $(0,\infty)$ reflected at 0, at least forward in time. 
Since the Wiener process is time-reversible, the formula holds also backward in time. However, even 
forward in time we could find no reference which 
gave the result \eqref{loctim-pdf} for general values of diffusivity $\kappa>0.$ Thus, we here very 
succinctly review the standard arguments which lead to  \eqref{loctim-pdf}, with references 
to more detailed discussions.  

If $\tilde{B}(t)$ is a standard Brownian motion (Wiener process), then the Brownian motion with diffusivity $\kappa$ 
started at $x\geq 0$ and reflected at 0 is given by the simple formula
 \begin{equation}\tilde{X}(t) =|\sqrt{2\kappa}\tilde{B}(t)-x|.  \end{equation}
Note that $\tilde{B}(t)$ has units of $({\rm time})^{1/2}$ whereas $\tilde{X}(t)$ has units of (length). 
The evolution of $\tilde{X}(t)$ can be obtained from the Tanaka equation for $a\geq 0:$  
\begin{eqnarray}
|\tilde{B}(t)-a|
                 &=& a+ \int_0^t {\rm sign}(\tilde{B}(t)-a) \rmd\tilde{B}(t) + \int_0^t ds\ \delta(\tilde{B}(s)-a) \cr
                 &=& a+\tilde{W}(t) + \tilde{L}_t(a),
\end{eqnarray}        
where $\tilde{W}(t)=\int_0^t {\rm sign}(\tilde{B}(t)-a)\rmd\tilde{B}(t) $ is another standard Brownian motion and  
$\tilde{L}_t(a)$ is the local time of $\tilde{B}(t)$ at point $a.$
For these well-known results, see \cite{karatzas1991brownian}, section 3.6 and \cite{rogers2000diffusions}, sections IV.6.43-44. 
(Possibly confusingly, there are different normalizations of local time in the literature, as discussed by \cite{karatzas1991brownian}, 
Remark 3.6.4 and \cite{rogers2000diffusions}, Remark IV.6.43.12. We follow the conventions of the latter.) 
Setting $a=x/\sqrt{2\kappa}$ in the Tanaka equation and multiplying through by $\sqrt{2\kappa}$ gives 
\begin{eqnarray}
\tilde{X}(t) &=& x+ \sqrt{2\kappa}\tilde{W}(t) + \kappa \int_0^t ds\ \delta(\tilde{X}(s)) \cr
                 &=& x+ \sqrt{2\kappa}\tilde{W}(t) + \kappa \tell_{0,t}(x).
\end{eqnarray}          
We used here that $\delta(\tilde{X}(t)-\epsilon)
=\delta(\sqrt{2\kappa}\tilde{B}(t)-x-\epsilon)+\delta(\sqrt{2\kappa}\tilde{B}(t)-x+\epsilon)$ for any $\epsilon>0$, in order
to relate the local times of $\tilde{B}(t)$ and $\tilde{X}(t)$. See \cite{rogers2000diffusions}, p.101 and \cite{karatzas1991brownian}, 
Remark 3.7.4. We have also used our normalization convention for the boundary local time density $\tell_{0,t}(x)$, which does not 
incorporate the factor of $\kappa$ into the definition of the local time density.

We next note the expression for the boundary local time density given by  the Skorohod equation 
\begin{equation}  \kappa \tell_{0,t}(x) = \max\{0, \sup_{0\leq s\leq t}(-x-\sqrt{2\kappa}\tilde{W}(s))\}. \label{skorohod} \end{equation}
See \cite{karatzas1991brownian}, Lemma 3.6.14. Define $\tilde{W}_-(t)=-\tilde{W}(t),$ which is another
standard Brownian motion, and denote its maximum up to time $t$ by  
$\tilde{M}(t)=\sup_{0\leq s\leq t} \tilde{W}_-(s), $ so that  
\begin{equation}
 \kappa \tell_{0,t}(x) = \max\{0, -x+\sqrt{2\kappa}\tilde{M}(t)\}. \label{skorohod-eq} \end{equation} 
It is well-known (and easy to see) that if $\tilde{\tau}(z)$ is the time for $\tilde{W}_-(t)$ to first hit $z,$ then 
 \begin{equation} P(\tilde{M}(t)>z) = P(\tilde{\tau}(z)<t).  \end{equation}
The latter probability can be evaluated by the reflection principle:  
\begin{equation} P(\tilde{\tau}(z)<t)=2 P(\tilde{W}_-(t)>z). \label{refl-princ} \end{equation}
See \cite{karatzas1991brownian}, section 2.6.  Thus, for $z>0,$
 \begin{equation}P(\tilde{M}(t)<z) = 2 P(\tilde{W}_-(t)<z)-1 = 2\Phi_t(z)-1 \end{equation}
with 
 \begin{equation} \Phi_t(z) = \frac{1}{\sqrt{2\pi t}}\int_{-\infty}^z dy\ \exp(-y^2/2t).  \end{equation}
Using this result and eq.(\ref{skorohod-eq}) gives
 \begin{equation} P(\tell_{0,t}(x)<\ell)=\left\{\begin{array}{cc}
                                      2\Phi_t\left(\frac{x+\kappa\ell}{\sqrt{2\kappa}}\right)-1 & \mbox{ for $\ell>0$}\cr
                                      0 & \mbox{ for $\ell<0$} 
                                      \end{array} \right. \end{equation}
Finally, differentiating with respect to $\ell$ gives \eqref{loctim-pdf} in the main text.

%%%%%%%%%%%%%%%%

\section{{ Heat Conduction with Oscillating Temperature B.C.}} \label{conductionFixedT}

{ We consider here consider the heat equation on the interval $[0,H]$ with an oscillating temperature imposed 
at the left endpoint. We work out explicit analytical results for both stochastic representations \eqref{hittingTimeRep}-\eqref{altDirFK}
and \eqref{theta-Dbc}, and briefly discuss the fluctuation-dissipation relation \eqref{NeumannVar} in the limit $\kappa\to 0$ 
at any fixed time $t.$ }

{ The problem considered is stated precisely as:
\begin{align}\nonumber
\partial_t T &= \kappa \partial_x^2 T\ \ \ \ \  {\rm for} \ \ \  x\in[0,H],\\ \label{pureCondfixedT}
T &= \sin(\omega t) \ \ \ \   {\rm at} \ \ \ x=0,\\ 
 T& = 0 \ \ \ \ \ \  \ \ \  \ \ \ {\rm at} \ \ \ x=H,\nonumber\\
 T& = 0 \ \ \ \ \ \  \ \ \ \ \  \ {\rm at} \ \ \ t=0.\nonumber
\end{align}
Here $T$ is made dimensionless by division with $\psi_0,$ so that the maximum value at the boundary $x=0$ is unity. 
The Green's function for the problem \eqref{pureCondfixedT} or, equivalently, the transition probability density 
of killed Brownian motion (absorbing boundaries), satisfies, for all $t>0$ and $x\in [0,H]$, the backward evolution equation:
\begin{align}\nonumber
\partial_s k(a,s|x,t) &=- \kappa \partial_a^2 k(a,s|x,t)\ \ \  {\rm for} \ \ \  (a,s)\in[0,H]\times [0,t]\\
 k(a,s|x,t) &= 0 \ \ \  {\rm at} \ \ \ a=0,H \label{pureCondfixedTTrans}\\
k(a,t|x,t) &=  \delta(a-x).\nonumber
\end{align}
The solution to \eqref{pureCondfixedTTrans} can be represented using Fourier sine series:
\begin{align}\label{fixedTsum1}
k(a,s; x,t) &=  \frac{2}{H} \sum_{n=1}^\infty \sin\left(\frac{n\pi x}{H}\right)\sin\left(\frac{n\pi a}{H}\right)e^{-\kappa\left(\frac{n \pi }{L}\right)^2(t-s)}\\  \label{fixedTsum2}
   &= \frac{1}{\sqrt{4\pi\kappa t}}\sum_{n=-\infty}^\infty  \left(e^{-\frac{ (x-a + 2n H )^2}{4\kappa t}}- e^{-\frac{(x+a + 2n H )^2}{4\kappa t}}\right).
\end{align}
The sum \eqref{fixedTsum1} converges rapidly for large $|t-s|$ whereas the second \eqref{fixedTsum2}, obtained by an application of the Poisson summation formula, converges rapidly for small $\kappa$.  We employ the latter form for subsequent analysis in this section. }

\subsection{{ Temperature Profile and Boundary Layer}}

{ According to the Feynman-Kac formula  \eqref{hittingTimeRep}-\eqref{altDirFK}, the temperature is represented by:
\begin{align} \label{fixedTtemp}
T(x,t) &= \kappa \int_0^t \!\!\! ds\  \sin(\omega s)  \left[\frac{\partial}{\partial a} k(a,s; x,t)\right]_{a=0},
\end{align}
where the probability flux through the boundary
\begin{align} \label{hittingTimeDist}
\kappa\left[\frac{\partial}{\partial a} k(a,s; x,t)\right]_{a=0}  &=  \frac{1}{\sqrt{4\pi\kappa(t-s)^3}}\sum_{n=-\infty}^\infty  (x+2nL) e^{-\frac{ (x+ 2n L )^2}{4\kappa (t-s)}}.
\end{align}
This can be interpreted as the probability distribution function for the first hitting time at the left boundary of Brownian motion on the finite interval $[0,H]$ starting at $(x,t)$ (see \cite{hsu1986brownian}).  As $\kappa\to 0$ at fixed $t$, all terms in \eqref{hittingTimeDist} vanish transcendentally except:
\begin{align}\label{hittingTimeDistHL}
\kappa\left[\frac{\partial}{\partial a} k(a,s; x,t)\right]_{a=0}   \sim  \frac{x}{\sqrt{4\pi \kappa(t-s)^3} } e^{-\frac{ x  ^2}{4\kappa (t-s)}}  \ \  \text{ as }  \ \   \kappa \to 0. 
\end{align}
Formula \eqref{hittingTimeDistHL} is identified as the first boundary hitting-time density function for Brownian motion on the half-line 
$[0,\infty)$, Eqn. \eqref{hittingtimePDF}.     
This well-known result (\cite{karatzas1991brownian}, section 2.6) 
can be obtained by differentiating $P(\tilde{\tau}(z) <t)$ in Eq.\eqref{refl-princ} with respect to $t$
and setting $t\to t-s.$ As a result, the temperature is given by:
\begin{align}\label{asymFFtemp}
T(x,t) &\sim \frac{x}{\sqrt{4\pi \kappa}}\int_0^t \!\!\! ds\  \frac{\sin(\omega s) }{ (t-s)^{3/2}}  
  e^{-\frac{x^2}{4\kappa (t-s)}}\ \  \text{ as }  \ \   \kappa \to 0.  
\end{align}
This is the exact solution to the heat equation on the half-line $[0,\infty)$ with the oscillating b.c. at $x=0$ in \ref{pureCondfixedT}. 
%Changing variables to $\mu = {x}/{\sqrt{4\kappa (t-s)}}$ \ we have:
%\begin{align}\label{asymFFtemp2}
%T(x,t) &\sim \frac{2}{\sqrt{\pi}}\int_{\frac{x}{\sqrt{4\kappa t}}}^\infty \! \!\!\! \!\!\! d\mu\ \  \sin\left(\omega\left( t- \frac{x^2}{4\kappa \mu^2}\right)\right) e^{-\mu^2}.
%\end{align}}
%{
%The expression (\ref{asymFFtemp2}) makes it apparent that the solution satisfies the boundary condition at $x=0$.  
It follows from (\ref{asymFFtemp})  that, with $x,t$ fixed, $T(x,t)\rightarrow 0$ as $\kappa\rightarrow 0.$ In that 
limit the temperature field is non-vanishing only in a thin boundary layer.}

{To analyze behavior in this layer, we introduce 
the scaling variable $\xi=x\sqrt{\omega/\kappa}$ and a dimensionless time $\beta=\omega(t-s)$ so that 
\begin{align}\label{asymFFtemp2}
T(\xi,t) &\sim \frac{\xi}{2\sqrt{\pi}}\int_0^{\omega t} \!\!\! d\beta\  \frac{\sin(\omega t-\beta) }{ \beta^{3/2}}  
  e^{-\xi^2/4\beta}. 
\end{align}
Since the integral (\ref{asymFFtemp2}) is absolutely convergent at $\beta=\infty$, we can divide it into two pieces,
$T(\xi,t)=T_{qs}(\xi,t)+T_{tr}(\xi,t),$ the first with the upper integration range extended to infinity 
\begin{equation}
 T_{qs}(\xi,t) =  \frac{\xi}{2\sqrt{\pi}}\int_0^{\infty} \!\!\! d\beta\  \frac{\sin(\omega t-\beta) }{ \beta^{3/2}}  
  e^{-\xi^2/4\beta}. 
\label{qs-T} \end{equation} 
and the complementary piece 
\begin{equation}
 T_{tr}(\xi,t) =  -\frac{\xi}{2\sqrt{\pi}}\int_{\omega t}^{\infty} \!\!\! d\beta\  \frac{\sin(\omega t-\beta) }{ \beta^{3/2}}  
  e^{-\xi^2/4\beta}, 
\label{tr-T} \end{equation}  
which tends to zero as $t\rightarrow \infty.$ The first term (\ref{qs-T}) represents quasi-steady conduction with long-time 
oscillation of the temperature at the frequency $\omega$ imposed by the boundary conditions, while the second term
(\ref{tr-T}) describes the transient heating-up phase that results from a uniformly cold, semi-infinite ``rod'' 
being exposed suddenly to a periodic temperature variation at its end. The quasi-stationary part can be evaluated 
explicitly by a standard Laplace transform (\cite{bateman1954tables}, 4.5(28), p. 146)
$$ \int_0^\infty \!\! d\beta \ \beta^{-3/2} \exp\left(-\sigma\beta-\alpha/4\beta\right)=
2\left(\frac{\pi}{\alpha}\right)^{1/2}\exp(-\sqrt{\sigma\alpha}), \quad {\rm Re}(\sigma)\geq 0, \,\,\,\, {\rm Re}(\alpha)>0, $$ 
implying, with $\alpha=\xi^2,$ $\sigma=i ,$ and $\sqrt{i}=(1+i)/\sqrt{2},$ that 
\begin{equation}
T_{qs}(\xi,t)=e^{-\xi/\sqrt{2}}\sin\left(\omega t - \xi/\sqrt{2} \right).
\label{qs-temp} \end{equation}}.

{The thermal dissipation in the scaled variables is given by 
\begin{eqnarray}\label{SolAsymGlobalDissFixedT}
 \langle \varepsilon_T \rangle_{V,t}&\equiv& \frac{1}{t} \int_0^tds \ \frac{1}{H}\int_0^Hdx\ \kappa|\partial_x T(x,s)|^2 \cr
 &=& \sqrt{\omega\kappa} \cdot \frac{1}{t} \int_0^tds\ \frac{1}{H}\int_0^{H(\omega/\kappa)^{1/2}} \!\!\!\!\!\! d\xi \,\,\, |\partial_\xi T(\xi,s)|^2. 
 \end{eqnarray}
In the limit $\kappa\to 0,$ the upper range $H(\omega/\kappa)^{1/2}$ of the $\xi$-integral can be set to infinity, since the integrand
decays rapidly to zero as $\xi\to\infty$ . This is equivalent to the statement that the thermal dissipation occurs entirely within the narrow 
temperature boundary layer of thickness $\sim (\kappa/\omega)^{1/2}.$ The quasi-steady contribution due to (\ref{qs-temp}) is found from 
\begin{equation}
\partial_\xi T_{qs}(\xi,t)=-e^{-\xi/\sqrt{2}}\sin\left(\omega t - \xi/\sqrt{2} +\frac{\pi}{4}\right)
\end{equation} 
and
\begin{equation}
\frac{1}{t} \int_0^tds\ \int_0^{\infty} \!\!\!\!\!\! d\xi \,\,\, |\partial_\xi T_{qs}(\xi,s)|^2
= \frac{1}{2\sqrt{2}}-\frac{\cos(2\omega t)-1}{2\omega t}\frac{1}{4\sqrt{2}}-\frac{\sin(2\omega t)}{2\omega t}\frac{1}{4\sqrt{2}}. \end{equation}
To evaluate the contribution from the transient, we use the derivative of (\ref{tr-T}) 
\begin{equation}
 \partial_\xi T_{tr}(\xi,t) =  -\frac{1}{2\sqrt{\pi}}\int_{\omega t}^{\infty} \!\!\! d\beta\  \frac{\sin(\omega t-\beta) }{ \beta^{3/2}}  
  \left(1-\frac{\xi^2}{2\beta}\right)e^{-\xi^2/4\beta}, 
\label{der-tr-T} \end{equation}   
which is again an absolutely convergent integral.  Substituting this integral and performing the Gaussian integrals over $\xi,$ 
one easily finds that
\begin{eqnarray}
\int_0^{\infty} \!\!\!\!\!\! d\xi \,\,\, |\partial_\xi T_{tr}(\xi,t)|^2&=&
\frac{3}{4}\sqrt{\pi}\int_{\omega t}^\infty d\beta_1\ \sin(\omega t-\beta_1)\int_{\omega t}^\infty d\beta_2\ \sin(\omega t-\beta_2)
\frac{1}{(\beta_1+\beta_2)^{5/2}}\cr
&\leq & \frac{3}{4} \left(\frac{\pi}{\omega t}\right)^{1/2}\int_1^\infty du_1 \int_1^\infty du_2 \frac{1}{(u_1+u_2)^{5/2}}=\left(\frac{\pi}{2\omega t}\right)^{1/2},
\end{eqnarray}
which provides an $O(1/\sqrt{\omega t})$ correction to the quasi-stationary result. Likewise, the cross-term between the quasi-stationary
and transient parts is 
\begin{eqnarray}
&& 2\int_0^{\infty} \!\!\!\!\!\! d\xi \,\,\, \partial_\xi T_{qs}(\xi,t)\partial_\xi T_{tr}(\xi,t) \cr
&& \hspace{30pt} = \frac{1}{\sqrt{\pi}} \int_{\omega t}^\infty \frac{d\beta}{\beta^{3/2}}\sin(\omega t-\beta) 
\int_0^\infty d\xi \ \sin\left(\omega t-\frac{\xi}{\sqrt{2}}+\frac{\pi}{4}\right) \left(1-\frac{\xi^2}{2\beta}\right) e^{-\xi/\sqrt{2}-\xi^2/4\beta} \nonumber
\end{eqnarray}
which is bounded by 
\begin{eqnarray*}
\left|2\int_0^{\infty} \!\!\!\!\!\! d\xi \,\,\, \partial_\xi T_{qs}(\xi,t)\partial_\xi T_{tr}(\xi,t)\right| 
&\leq& \frac{1}{\sqrt{\pi}} \int_{\omega t}^\infty \frac{d\beta}{\beta^{3/2}} 
\int_0^\infty d\xi \ \left(1+\frac{\xi^2}{2\beta}\right) e^{-\xi/\sqrt{2}} \cr
&=&2\left(\frac{2}{\pi \omega t}\right)^{1/2} \left(1+ \frac{2}{3\omega t}\right)
\end{eqnarray*}
Gathering all these results together,
$$ \langle \varepsilon_T \rangle_{V,t} = \frac{\psi_0^2\sqrt{\omega\kappa}}{2\sqrt{2}H}\left[1+O\left(\frac{1}{\sqrt{\omega t}}\right)\right], $$
with the leading contribution coming from time-average of the quasi-steady term. Here we have restored the factor $\psi_0^2$ from the amplitude
of the oscillating boundary temperature.}
 
\begin{figure}
\centering
	\includegraphics[width=0.7\linewidth,height=0.45\linewidth]{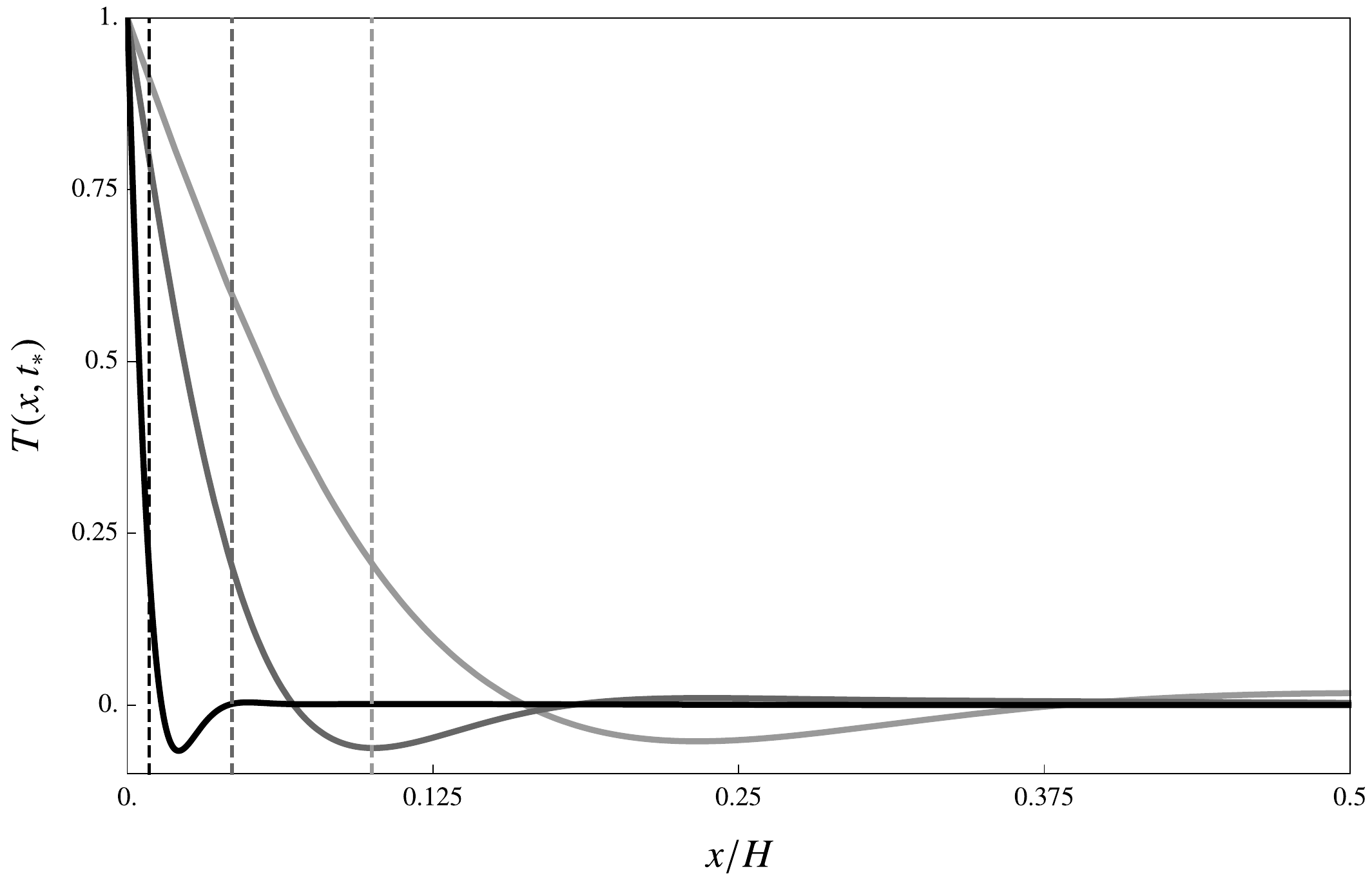}
	\caption{{Temperature distribution for pure conduction on a 1D interval with oscillating Dirichlet boundary conditions, depicted at 
	a fixed instant $t_*$ such that $t_*\sqrt{\omega \kappa}/H=1$.
	Plotted are solutions for three pairs of $(\omega,\kappa)$ satisfying $(\omega  t_*)(\kappa t_*/H^2)= 1$; these are $\omega  t_* = 9\frac{\pi}{2}$ (light grey), $\omega  t_* = 21\frac{\pi}{2}$ (grey) and  $\omega  t_* = 101\frac{\pi}{2}$ (black).  The dashed vertical lines, which indicate the boundary layer region, are placed at $(2\kappa/\omega)^{1/2}$.}}
\label{fig:TempProfile}
\end{figure}

{Apart from the numerical pre-factor, this result could have been expected on dimensional grounds; see eq.\eqref{dimAnalFixedT} of Section \ref{sec:SS-FDR-Dbc}.
In order to obtain a result non-vanishing in the limit $\kappa\to 0,$ the frequency $\omega$ must go to infinity and, in that case, 
the time-dependent corrections all vanish.  Assuming that $\omega\sim \kappa^{-\alpha}$ as $\kappa\to 0$ , we have: 
\begin{align*}
0\leq \alpha &< 1 \quad \text{ then } \quad \lim_{\kappa \to 0}   \langle \varepsilon_T \rangle_{V,t}=0,\\
\alpha &= 1 \quad \text{ then } \quad \lim_{\kappa \to 0}   \langle \varepsilon_T \rangle_{V,t}=({\rm const.}),\\
\alpha &> 1 \quad \text{ then } \quad \lim_{\kappa \to 0}   \langle \varepsilon_T \rangle_{V,t}=+\infty.
\end{align*}
Note that $\omega\sim \kappa^{-1}$  illustrates the scenario described in Section \ref{sec:SS-FDR-Dbc}  since the temperature gradients have magnitude $\partial_x T\sim 1/\kappa$ in a boundary layer of thickness $(\kappa/\omega)^{1/2}\sim \kappa$. See Figure \ref{fig:TempProfile} for plot of steepening profile \eqref{asymFFtemp}, indicating temperature gradients growing unboundedly large in the boundary layer as $\kappa\to 0$, $\omega=\kappa^{-1}\to \infty$.}

\subsection{{ Alternative Feynman-Kac Formula and Fluctuation-Dissipation Relation}} \label{altFeynKacAppend}

{In section \ref{FDR-equal-Dbc} of the main body of the text, we showed that there are two equivalent stochastic 
representations valid for scalars supplied with Dirichlet boundary conditions. One of these representations involves 
first-hitting times \eqref{altDirFK} and the other involves local-time integrals \eqref{theta-Dbc}.  The utility of the former 
representation \eqref{altDirFK} was demonstrated in the previous subsection where it was used to explicitly solve 
the heat equation \eqref{asymFFtemp2}.  The latter formula \eqref{theta-Dbc}  is also theoretically useful as it allows 
us to obtain an FDR equality for Dirichlet boundary conditions (see Eq. \eqref{NeumannVar} of Section \ref{FDR-equal-Dbc}). 
On the other hand, the formula \eqref{theta-Dbc} is computationally unwieldy since it requires sampling the (a priori unknown) 
flux $g(x,t) := \kappa \partial_x T(x,t)$ at the boundary with \emph{reflecting} (not killed) Brownian motion.  Here, as a 
consistency check, we demonstrate the equivalence of these formulae directly for conduction on the semi-infinite space 
interval $(0,\infty)$ with oscillating boundary condition at $x=0.$} %and calculate the dissipation from our exact FDR. }

{A formula for the flux follows by direct differentiation of   \eqref{asymFFtemp}, which yields
\begin{align}
\nonumber 
g(x,t) &= - \sqrt{\frac{\kappa}{\pi}} \int_{0}^t \!\! ds\  \frac{\sin(\omega s) }{(t-s)^{3/2}}   \frac{1}{2}\left(1-\frac{x^2}{2\kappa (t-s)}\right) e^{-\frac{x^2}{4\kappa (t-s)}} \\
\nonumber
&= - \sqrt{\frac{\kappa}{\pi}} \int_{0}^t \!\! ds\  \sin(\omega s) \frac{\partial}{\partial s} \left[ \frac{e^{-\frac{x^2}{4\kappa (t-s)}}}{(t-s)^{1/2}}\right]\\
\label{fixedTDtemp}
&= \omega \sqrt{\frac{\kappa}{\pi}}\int_{0}^t \!\! ds\  \frac{\cos(\omega s) }{(t-s)^{1/2}}   e^{-\frac{x^2}{4\kappa (t-s)}},
\end{align}
after an integration by parts. Thus, the temperature profile according to the local-time representation \eqref{theta-Dbc} is given by 
\begin{align}\label{altTempRep}
T(x,t) = \int_{0}^t \!\! ds \ g(0,s)  p(0,s|x,t) = \frac{\omega}{\pi}   \int_{0}^t \!\! ds  \int_{0}^s \!\! ds'  \frac{\cos(\omega s')}{\sqrt{s-s'}\sqrt{t-s}}   e^{-\frac{x^2}{4\kappa(t-s)}},
 \end{align}
where we have used the transition density for reflecting Brownian motion  \eqref{zeroDiffasymTrans2}. 
Interchanging the order of the integrals, 
\begin{align}\label{altTempRep}
T(x,t) = \frac{\omega}{\pi}   \int_{0}^t \!\! ds' \cos(\omega s') \int_{s'}^t \!\! ds  
\frac{e^{-\frac{x^2}{4\kappa(t-s)}}}{\sqrt{s-s'}\sqrt{t-s}}. 
 \end{align}
Under the substitutions $u=x^2/4\kappa(t-s)$ and $b=x^2/4\kappa(t-s')$ the inner integral becomes
$$  \int_{s'}^t \!\! ds  \frac{e^{-\frac{x^2}{4\kappa(t-s)}}}{\sqrt{s-s'}\sqrt{t-s}}=\sqrt{b}\int_b^\infty du \frac{e^{-u}}{(u-b)^{1/2}u}
=\pi\ {\rm erfc}(\sqrt{b}),
$$   
in terms of the complementary error function ${\rm erfc}(z),$ by \cite{bateman1954tables}, 4.2(26), p. 136. 
%\begin{align}\label{explicitTransDens}
%k(0,s|x,t) \sim\frac{1}{\sqrt{\pi \kappa (t-s)}}e^{-\frac{x^2}{4\kappa(t-s)}}\ \  \text{ as }  \ \   \kappa \to 0.  
% \end{align}
%Substituting \eqref{explicitFlux} and \eqref{explicitTransDens} into the formula \eqref{altTempRep}, we obtain the following explicit representation for the temperature distribution:
%  \begin{align}\label{altTempRep2}
%T(x,t) \sim \frac{\omega}{\pi}   \int_{0}^t \!\! ds  \int_{0}^s \!\! ds'  \frac{\cos(\omega s')}{\sqrt{s-s'}\sqrt{t-s}}   e^{-\frac{x^2}{4\kappa(t-s)}}
% \end{align}
%First, note that $T(0,t)=\sin(\omega t)$ can be seen directly from \eqref{altTempRep} by exchanging the order of integration.  To see that the two representations \eqref{asymFFtemp} and \eqref{altTempRep} are equivalent, exchange the orders of integration in  {\eqref{altTempRep}} and changing variables to $\mu = {x}/{\sqrt{4\kappa (t-s)}}$:
The local-time representation formula \eqref{altTempRep} therefore gives 
\begin{align}
T(x,t) 
%&\sim \frac{\omega}{\pi\sqrt{\kappa}}   \int_{0}^t \!\! ds' \cos(\omega s')  \int_{\frac{x}{\sqrt{4\pi(t-s')}}}^\infty \!\! \!\! \!\! d\mu  \  x\mu^{-2} \left(t-s'- \frac{x^2}{4\kappa \mu^2}\right)^{-1/2}   e^{-\mu^2}\\ \nonumber
%&=  \frac{4 \omega\sqrt{\kappa} }{\pi}   \frac{d}{dx} \int_{0}^t \!\! ds' \cos(\omega s')  \int_{\frac{x}{\sqrt{4\pi(t-s')}}}^\infty \!\! \!\! \!\! d\mu  \ \left(t-s'- \frac{x^2}{4\kappa \mu^2}\right)^{1/2}   e^{-\mu^2}\\ 
&= \omega \int_{0}^t \!\! ds' \ \cos(\omega s') \ {\rm erfc}\left(\frac{x}{\sqrt{4\kappa(t-s')}}\right).\label{altTempRep2}
 \end{align}
 Using the expression \eqref{altTempRep2} and $\omega\cos(\omega s')=(\partial/\partial s')\sin(\omega s'),$ a straightforward 
 integration by parts in $s'$ recovers the formula  \eqref{asymFFtemp} from the hitting-time representation. Thus, the two 
 stochastic representations are verified to yield identical results for this problem.}
 
{ Additionally, as we did in section \ref{checkNeumannFDR} for the case of temperature supplied with Neumann conditions, 
one can verify directly that the second form \eqref{NeumannVar} of our fluctuation-dissipation relation equals the total dissipation 
\eqref{SolAsymGlobalDissFixedT}.  The FDR \eqref{NeumannVar} for the present problem on the space interval $(0,H)$ reduces 
asymptotically as $\kappa\to 0$ to:
\begin{align*}
 \langle\varepsilon_T^{fluc}(x)\rangle_{V,t}&= \frac{1}{H}\int_0^H\!\!\! dx \ \frac{1}{2} {\rm Var}\left[ \int_{0}^t { \kappa\partial_xT(\txi_{t,s}(x),s)   \ \hd \tell^L_{t,s}}(x) \right]
   \\
   &=  \frac{1}{H}\int_0^H\!\!\! dx \int_{0}^tds\int_{s}^tds'  \ g(0,s)g(0,s')  \left[p(0,s|0,s')-p(0,s|x,t)\right]p(0,s'|x,t),
\end{align*} 
where, again, the fluxes $g(0,t)$ are explicitly defined by \eqref{fixedTDtemp}. % Asymptotically the contribution from the transition densities reduces to:
%\begin{align*}
%\sqrt{\pi\kappa} \int_0^H\!\!\! dx \left[p(0,s|0,s')-p(0,s|x,t)\right]p(0,s'|x,t) \sim\frac{1}{\sqrt{s'-s}}- \frac{1}{\sqrt{t-s+t-s'}}\ \  \text{ as }  \ \   \kappa \to 0.  
% \end{align*}
 %The dissipation becomes, with $y(t)= 2t\omega$:
 %\begin{align*}
 %\langle\varepsilon_T^{fluc}(x)\rangle= 2\kappa \omega&\int_{0}^tds\int_{s}^tds'  \left(\frac{1}{\sqrt{s'-s}}- \frac{1}{\sqrt{t-s+t-s'}}\right)\times\\
 %& \left(C(\sqrt{y(s)/\pi})\cos(y(s)/2)+S(\sqrt{y(s)/\pi})\sin(y(s)/2)\right) \times\\
 %&\left(C(\sqrt{y(s')/\pi})\cos(y(s')/2)+S(\sqrt{y(s')/\pi})\sin(y(s')/2)\right)
%\end{align*}
A straightforward but lengthy calculation, similar to the one we performed to show the equivalence of the two Feynman-Kac representations, 
verifies that one obtains the same asymptotic expression for the dissipation \eqref{SolAsymGlobalDissFixedT} directly from our exact fluctuation-dissipation relation.}

%\section{Mathematical Proofs}\label{proofs}

%Here we give the details of the proofs in the main text. 

\section{Spontaneous Stochasticity Implies Anomalous Scalar Dissipation
for Dirichlet Scalar B. C. }\label{App:A3}

{We supply here details of the proofs in section \ref{sec:SS-FDR-Dbc} that 
spontaneous stochasticity implies anomalous dissipation of a passive scalar 
with appropriate choices of initial data, Dirichlet boundary data, and bulk scalar sources.}  
For active scalars, it seems likely that this remains true but, due to lack of freedom in choosing 
the scalar data, we cannot rigorously draw such conclusions.

\subsection{Vanishing scalar sources} 

For zero scalar sources, we exactly mimic the arguments of Appendix A of {paper I}
with virtually no change to show that,  
if there is spontaneous stochasticity, so that non-factorization occurs in (\ref{q2-fac}) for a set of $\bx$
of positive measure, then there exists a smooth function $\Theta=(\theta_0,\psi)$ on $\cS$ which makes the 
variance \eqref{Varbound} positive. This provides a positive lower bound to the global cumulative dissipation 
with that choice of $\Theta$ for a suitable subsequence $\nu_j,\kappa_j\rightarrow 0$.

Unfortunately, it seems to be very difficult to show that, with boundary conditions $\psi$ fixed, spontaneous stochasticity implies 
anomalous dissipation for a suitable choice of initial data $\theta_0$. The difficulties are made apparent by the following easily 
proved identity for the variance \eqref{Varbound}, which separates contributions from $\theta_0$ and $\psi$:
\begin{align}\nonumber
& \var\left[\Theta(\tilde{\bsigma}(\bx,t)) \right]  
= \var\left[\theta_0(\tbxi_{t,0}(\bx))\Big| \tilde{\tau}(\bx,t)=0\right] P\big(\tilde{\tau}(\bx,t)=0\big) \\
& \hspace{50pt}   +\var\left[\psi(\tbxi_{t,\tilde{\tau}(\bx,t)}(\bx),\tilde{\tau}(\bx,t))\Big| \tilde{\tau}(\bx,t)>0\right] P\big(\tilde{\tau}(\bx,t)>0\big) .  \label{ICandBDVar}
\end{align}
Here the variances are conditioned upon the events $\tilde{\tau}(\bx,t)=0$ or $\tilde{\tau}(\bx,t)>0$ and the 
symbol $P(\cdots)$ denotes the probability measure of these same events.  By non-negativity of the variance, 
one gets readily 
\begin{equation}  \var\left[\Theta(\tilde{\bsigma}(\bx,t)) \right]  
\geq \var\left[\theta_0(\tbxi_{t,0}(\bx))\Big| \tilde{\tau}(\bx,t)=0\right] P\big(\tilde{\tau}(\bx,t)=0\big). \end{equation}
If one assumes that there is spontaneous stochasticity for the sub-ensembles specified by $\tilde{\tau}(\bx,t)=0$
for a positive-measure set of $\bx\in \Omega$ (which is not an entirely straightforward assumption), then we can again 
use the arguments of {Paper I, Appendix A} to prove that  
\begin{equation}  \lim_{j\rightarrow\infty} \var\left[\theta_0(\tbxi_{t,0}^{\nu_j,\kappa_j}(\bx))\Big| \tilde{\tau}^{\nu_j,\kappa_j}(\bx,t)=0\right] >0 \end{equation}
for a suitable smooth initial datum $\theta_0$ and for a suitable subsequence $\nu_j,\kappa_j\rightarrow 0,$ on a positive measure 
set of $\bx\in \Omega.$  However, two difficulties appear. One is that, possibly,  
\begin{equation}   \lim_{j\rightarrow\infty} P\big(\tilde{\tau}^{\nu_j,\kappa_j}(\bx,t)=0\big)=0\quad a.e. \ \bx\in \Omega\ !   \end{equation}
Another technical difficulty is that vague topology of Young measures can be used to establish convergence of the space-average 
\begin{equation}  \lim_{j\rightarrow\infty} \left\langle
\var\left[\theta_0(\tbxi_{t,0}^{\nu_j,\kappa_j})\Big| \tilde{\tau}^{\nu_j,\kappa_j}(t)=0\right]\right\rangle_\Omega  \end{equation}
whenever $\theta_0$ is a continuous function, but to obtain convergence of the needed space-average 
\begin{equation}  \lim_{j\rightarrow\infty} \left\langle
\var\left[\theta_0(\tbxi_{t,0}^{\nu_j,\kappa_j})\Big| \tilde{\tau}^{\nu_j,\kappa_j}(t)=0\right]
P\big(\tilde{\tau}^{\nu_j,\kappa_j}(t)=0\big)
\right\rangle_\Omega  \end{equation}
by using the same arguments, one must know that the functions 
\begin{equation}  g^{\nu,\kappa}(\bx,t) \equiv P\big(\tilde{\tau}^{\nu,\kappa}(\bx,t)=0\big)  \end{equation}
converge as $\nu,\kappa\rightarrow 0$ to a continuous (non-zero) function $g(\bx,t)$ uniformly in $\bx.$ This is a non-trivial
statement which cannot be obtained by mere compactness arguments.  

\subsection{Non-vanishing scalar sources}

To deal with scalar sources, we must introduce a new type of transition probability 
\begin{equation}  q^{\nu,\kappa}(\tau,\by,s|\bx,t)=\bE\left[ \delta(\tau-\tilde{\tau}(\bx,t)) \delta^d(\by-\tbxi_{t,s}(\bx)) \right],
\quad \by\in \Omega,\ 0\leq \tau<s<t \end{equation}
which gives the joint probability for the particle released at $\bx$ at time $t$ to first hit the space boundary $\partial\Omega$ at time $\tau$
(going backward) and also to pass through point $\by$ at time $s$ satisfying $\tau<s<t$. We can then rewrite the 
stochastic representation (\ref{DirFK}) as
\begin{equation}  \theta(\bx,t) = \int_\cS H^d(d\bsigma)\  \Theta(\bsigma)\  q(\bsigma|\bx,t) + 
\int_0^t\rmd\tau \ \int_\tau^t \rmd s\ \int_\Omega d^dy\ S(\by,s) q(\tau,\by,s|\bx,t). \end{equation}
To obtain a similar expression for the source variance term, we must introduce corresponding 2-time transition densities
\begin{eqnarray}
&& q^{\nu,\kappa}_2(\tau; \by,s;\tau',\by',s'|\bx,t)= \cr
&& \hspace{75pt} \bE\left[ \delta(\tau-\tilde{\tau}(\bx,t)) \delta^d(\by-\tbxi_{t,s}(\bx))  \delta(\tau'-\tilde{\tau}(\bx,t))  
\delta^d(\by'-\tbxi_{t,s'}(\bx))\right], \cr
&& \hspace{175pt} \by,\by'\in \Omega,\ \tau<s<t, \tau'<s'<t. 
\end{eqnarray}
and 
\begin{equation} q_2(\tau,\by,s;\tau',\by',s'|\bx,t)=\delta(\tau-\tau')q_2(\tau;\by,s;\by',s'|\bx,t).\end{equation}
With this quantity we can write 
\begin{eqnarray}
&& {\rm Var} \left[ \int_{\tilde{\tau}(\bx,t)}^t \rmd s\ S(\tbxi_{t,s}(\bx) ,s)   \right]  \cr
&& \hspace{30pt} =  \int_{0}^t d\tau \int_{0}^t d\tau'   \int d^dy\int d^dy' \int_{\tau}^t ds \int_{\tau'}^t ds' \ S(\by,s) S(\by',s') \times \cr
&& \hspace{60pt} \Big[q_2(\tau,\by,s;\tau',\by',s'|\bx,t)-q(\tau,\by,s|\bx,t)q(\tau',\by',s'|\bx,t)\Big].  
\label{VarS-Dbc} \end{eqnarray}
We can also write the initial condition-source covariance in a similar fashion, using the 2-time probability density
\begin{equation}  q^{\nu,\kappa}_2(\bsigma; \tau',\by',s'|\bx,t) =\bE\left[\delta_\cS\left(\bsigma,\tilde{\bsigma}(\bx,t)\right)
\delta(\tau'-\tilde{\tau}(\bx',t)) \delta^d(\by'-\tbxi_{t,s'}(\bx'))\right] , \end{equation}
which is proportional to $\delta(\tau-\tau'),$ if we write $\bsigma=(\by,\tau).$ We have 
\begin{align}\nonumber
{\rm Cov} &\left[\Theta(\tbxi_{t,\tilde{\tau}(\bx,t)}(\bx),\tilde{\tau}(\bx,t)), \int_{\tilde{\tau}(\bx,t)}^t \rmd t'\ 
S(\tbxi_{t,t'}(\bx) ,t')   \right]  \\ \nonumber
& \ \ \ \ =  \int_\cS H^d(d\bsigma) \int_{0}^t d\tau'  \int d^dy' \int_{\tau'}^t ds'\ \Theta(\by,s) S(\by',s') 
  \times\\
 & \ \  \ \ \ \ \ \ \ \   \Big[q_2(\bsigma; \tau',\by',s'|\bx,t)-q(\bsigma|\bx,t)q(\tau',\by',s'|\bx,t)\Big], 
\end{align}
as the corresponding representation. The very close similarity of these formulas to those derived previously for advected scalars 
in flows without boundaries and for scalars with insulating/adiabatic walls allows us to prove that spontaneous stochasticity 
is sufficient for anomalous dissipation of passive scalars also for Dirichlet b.c. in the presence of smooth sources. 
The existence of a smooth source $S$
on $\Omega\times [0,t]$ which makes the variance \eqref{VarS-Dbc} positive is likewise shown by arguments like 
those in Appendix A of {paper I}. We leave details to the reader.     The statement here is that a smooth source 
$S$ can be chosen for which there is anomalous scalar dissipation, either with 
zero initial-boundary data or else with a smooth non-zero choice $\Theta=(\theta_0,\psi)$ of such data.

\section{Numerical Methods}\label{numerics}
\subsection{Methods for Section \ref{StochRep-FDR-Nbc}}\label{channelAppendix}

As in our previous numerical study of isotropic turbulence, we solve the stochastic equations (\ref{stochflowreflected}),
(\ref{loctimden-form}) with interpolated velocities retrieved using the {\tt getVelocity} database function. 
To solve these backward It$\bar{{\rm o}}$ SDE's, we use a reflected time $\hat{s}=t_f-s$ which transforms 
them into forward It$\bar{{\rm o}}$  SDE's. 
To implement the reflecting boundary conditions at the channel walls we have found it convenient 
to use the very simple Euler algorithm of \cite{lepingle1995euler}, which is ideally suited to the 
parallel plane boundaries of the channel flow. This method reduces to the standard Euler-Maruyma scheme at some specified 
small distance $\epsilon$ away from the walls. This algorithm (dropping hats 
on the reflected time)  takes at the $k$th time-step 
\begin{equation} \tbxi(s_{k+1}) =  \tbxi(s_k)+ \Delta\tbxi(s_{k+1};s_k) + \kappa{\bf n}^+ \ \Delta\tell^+(s_k) 
+\kappa{\bf n}^- \ \Delta\tell^-(s_k) \label{LePing-xi} \end{equation}
where $\Delta\tbxi(s;s_k)$ is given by the Euler-Maruyama discretization
 \begin{equation} \Delta\tbxi(s;s_k)= \bu(\tbxi(s_k),s_k)(s-s_k) + \sqrt{2\kappa} (\bW(s)-\bW(s_k))\end{equation}
and where $\Delta\tell^\pm(s_k)=\tell^\pm(s_{k+1})-\tell^\pm(s_k)$ is given by the Skorohod equation (see 
\cite{karatzas1991brownian}, Lemma 3.6.14 and Eq.~\eqref{skorohod}):
 \begin{equation}
 \kappa \Delta\tell^\pm(s_k) = \max\Big\{0,\max_{s_k\leq s\leq s_{k+1}}
 \big\{-{\bf n}^\pm\cdot[\tbxi(s_k)+\Delta\tbxi(s;s_k)]\big\}-h\Big\}. 
 \label{LePing-ell} \end{equation} 
At the top wall $y=+h,\ {\bf n}^+=-\hat{{\bf y}}$ and at the bottom wall $y=-h,\ {\bf n}^-=+\hat{{\bf y}}.$ The stochastic solution 
of the Skorohod equation (\ref{LePing-ell}) is exactly realized by a closed formula involving an independent exponential random variable.
See \cite{lepingle1995euler}, Theorem 1. To minimize the additional computational cost of solving the Skorohod equation 
\eqref{LePing-ell}, the algorithm of \cite{lepingle1995euler} adds the local time density terms into Eq.~\eqref{LePing-xi} 
only when the particle is within a distance $\epsilon$ of the walls, fixed independent of $\Delta s=s_{k+1}-s_k$\footnote{In principle, 
the boundary local time densities for channel-flow that appear in Figs.~\ref{fig3},\ref{fig:channel} are thus a sum of contributions from 
both top and bottom walls. However, in practice, no particles released at the bottom wall were ever observed to be 
incident upon the top wall within the entire time available in the database and thus $\tell(s)=\tell^-(s)$ . This is to be expected, 
since wall-normal velocities in turbulent channel-flow generally have maximum magnitudes of order $u_\tau$ and, in the archived 
flow, advection at this maximum speed would require about double the archived time interval for a particle to transit from the bottom wall to the top wall.}.  
We chose $\epsilon=10\delta_\nu$ in our channel-flow simulations.

The \cite{lepingle1995euler} algorithm outlined above yields strong approximations to the joint process 
$(\tbxi_{t,s}(\bx),\tell_{t,s}(\bx))$ for any $\bx.$ 
To study numerically the strong convergence of the approximate solutions of Eqs.~\eqref{LePing-xi},\eqref{LePing-ell} for the 
results presented in Fig.~\ref{fig3}, we first generated a realization of a Brownian motion $\bW(s_i)$ sampled on an extremely  
fine grid of times $s_i$ with a very small time-step $\delta s.$ We then applied the above algorithm of \cite{lepingle1995euler} 
for a larger time-step $\Delta s=m \ \delta s$ which is a multiple of $\delta s$ by a big integer $m,$ using the same previously
sampled Brownian motion $\bW(s_i)$ for all $\Delta s=m \delta s$. We reduced the size of $m$ until the numerical 
results for the joint process $(\tbxi(s),\tell(s))$ were converged to within 0.1\%, which required a very small
time-step $\Delta s=6.5\times 10^{-6},$ or $1/1000$th of the time between the stored database frames. 
The requirements are much less stringent for weak convergence of statistical averages over Brownian motions than 
for the strong convergence at fixed $\bW_s$. The averages presented in section \ref{StochRep-FDR-Nbc}
were calculated with a time-step of $\Delta s =2.1\bar{6} \times10^{-3}$, or $1/3$ of the time between database frames. 
To check for weak convergence, we numerically integrated the equations
with double this time-step and found a relative change in the ensemble averages of less than 0.1\%.

The $N$-sample averages used for particle dispersions and PDF's were the same as 
those for isotropic turbulence discussed in Appendix {C of paper I}. In particular, local time PDFs were 
estimated with kernel density techniques. One key difference is that the absolute local times $|\tell_{t,0}|$ are positive 
quantities, so that their PDF's have a support boundary at 0 and naive application of kernel density estimators leads 
to ``seepage" of mass to negative values. Since the exact PDF is expected to have a maximum at 0 (highest probability 
for particles never to hit the wall), we employ the ``reflection method" \citep{karunamuni2005boundary}.  This procedure involves 
performing the kernel density estimation on a symmetric dataset constructed by augmenting the original data by its negation. 
The optimal bandwidth was again chosen by the ``principle of minimal sensitivity''. Using these procedures and the same number 
of samples ($N=1024$) as for channel-flow, we were able to recover the analytical results in Appendix \ref{reflBr} 
for the PDF of local times of a Brownian motion on the half-axis reflected at the origin. 
 
% \begin{figure}
%\centering    
%	\includegraphics[width=.7\linewidth,height=.7\linewidth]{MinimalSens_lt}
%	\caption{Minimal sensitivity for local time. Keep?}
%\label{fig:minimalSens}
%\end{figure}
  
\subsection{Methods for Section \ref{StochRep-FDR-Dbc}}\label{HittingTimeAppendix}

To obtain the hitting times for channel-flow from the numerical solution of the SDE \eqref{stochflowreflected} with the 
Euler-Maruyama scheme, we simply monitored the solution until the first time-step where the $y$-component satisfied
$\teta(s_k)>-h$ but $\teta(s_{k+1})<-h.$ We then stopped the integration and set $\tilde{\tau}=(s_k+s_{k+1})/2.$  
With a time-step of $\Delta s=2.1\bar{6}\times 10^{-3},$ this gives the hitting-time accurately to within $1.08\bar{3}\times 10^{-3},$
or 5.41\% of the viscous time $\tau_\nu$. To obtain PDFs of $\lambda=\ln((t_f-\tau)/\tau_\nu)$ from the channel flow data, we applied 
the same kernel density techniques as in Appendix {C of paper I}. Note for Brownian motion 
(see Eq.~\eqref{hittingtimePDF2} below) the mean hitting is infinite and  with fluid-advection the hitting 
times are expected to be even larger. Thus, many solutions do not ever hit the wall over the total time available 
in the channel-flow database and this necessitates a very large number of samples. The results presented in the text
were for N=14,336 samples. The PDF's were verified to be converged with $\Delta s=
2.1\bar{6}\times 10^{-3}$ to within 0.1\% and the plotted error bars represent both s.e.m. 
and the change due to $10\%$ increase in $h_*,$ as described in Appendix \ref{channelAppendix}.

\begin{figure}
\centering    
	\includegraphics[width=.9\linewidth,height=.4\linewidth]{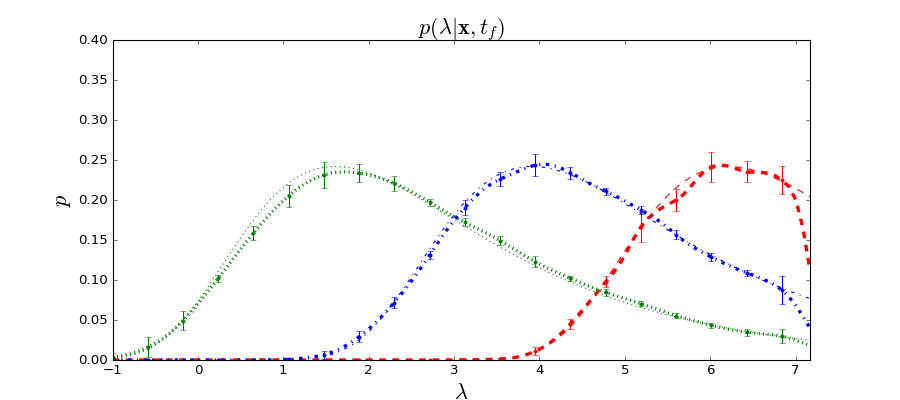}
	\caption{Hitting-time PDF's of Brownian motion 
	for $Pr = 0.1$ (\textcolor{ForestGreen}{green}, dot, \dottedrule), $1.0$ (\textcolor{blue}{blue}, dash-dot, \dasheddottedrule) and $10$ (\textcolor{red}{red}, dash, \dashedrule).
	Numerical results are plotted with heavy lines, analytical results with light lines.}
\label{fig:brownianHT}
\end{figure}

We applied identical numerical procedures to calculate  
the PDFs of $\lambda=\ln((t_f-\tau)/\tau_\nu)$ for a Brownian motion with diffusivity $\kappa$ started at position $y>-h$
at time $t=t_f$ and going backward in time to $t=0$ . {The analytical result to be used for comparison 
with channel flow is given by formula \eqref{hittingtimePDF} with $x\to \Delta y:=y+h>0,$ $\, t-s\to t_f-\tau$ }:  
\begin{align}\label{hittingtimePDF2}
p^\kappa(\tau|y,t_f) = 
\begin{cases} 
 \frac{\Delta y}{\sqrt{4\pi\kappa (t_f- \tau)^3}}\exp\left(-\frac{(\Delta y)^2}{4\kappa (t_f-\tau)}\right)   & \tau<t_f\\
0 & \tau > t_f 
\end{cases}.
\end{align}
Numerical and analytical results for Brownian motion are plotted together in Figure \ref{fig:brownianHT}. For 
$\lambda<1.6$ one can see that the error from time-discretization leads to a $\sim 5\%$ shift of the PDF to the 
right, toward larger hitting times. To correct this would require either a smaller time-step or a more sophisticated stochastic 
interpolation scheme.  For $\lambda>6.5,$ the lack of samples for $\lambda=\ln(t_f/\tau_\nu)\doteq 7.17$ leads the 
kernel density estimators to clearly under-predict the PDF's. Similar errors must exist in the channel-flow results.

\bibliographystyle{jfm}

\bibliography{bibliography}

\end{document}